\documentclass[10pt, final]{IEEEtran}

\usepackage{amsthm}

\usepackage{amssymb}
\usepackage{amsmath}
\usepackage{bbm}
\usepackage[ruled]{algorithm2e}
\usepackage{mathrsfs}
\usepackage{cite}
\usepackage{bm,epsfig,amsthm,url}
\usepackage{indentfirst}
\usepackage{amsfonts}
\usepackage{epstopdf}
\usepackage{cases}
\usepackage[caption=false,font=scriptsize]{subfig}
\usepackage{xcolor}
\usepackage{mathtools}

\makeatletter
\renewcommand{\maketag@@@}[1]{\hbox{\m@th\normalsize\normalfont#1}}%
\makeatother
\usepackage{hyperref}
\usepackage{stfloats}

\newtheoremstyle{mystyle}{}{}{}{}{}{: }{0pt}{\indent \it{\thmname{#1}\thmnumber{ #2}\thmnote{#3}}}
\theoremstyle{mystyle}

\newtheorem{Proposition}{Proposition}

\begin{document}
\title{Reconfigurable Codebook-Based Beamforming for RDARS-Aided mmWave MU-MIMO Systems}
\author{Chengwang~Ji, Qing~Xue, Haiquan~Lu, Jintao~Wang, Qiaoyan~Peng, Shaodan~Ma and~Wei~Zhang
\thanks{Chengwang Ji, Jintao Wang, Qiaoyan Peng, Haiquan Lu, and Shaodan Ma are with the State Key Laboratory of Internet of Things for Smart City and the Department of Electrical and Computer Engineering, University of Macau, Macao SAR, China (e-mails: \{ji.chengwang, wang.jintao, qiaoyan.peng\}@connect.um.edu.mo; haiq\_lu@163.com; shaodanma@um.edu.mo). 
Qing Xue is with the School of Communication and Information Engineering, Chongqing University of Posts and Telecommunications, Chongqing 400065, China (e-mail: xueq@cqupt.edu.cn).
Wei Zhang is with the School of Electrical Engineering and Telecommunications, University of New South Wales, Sydney, NSW 2052, Australia (e-mail: w.zhang@unsw.edu.au).}
\vspace{-25pt}
}
%


%
\maketitle

\begin{abstract}
Reconfigurable distributed antenna and reflecting surface (RDARS) is a new architecture for the sixth-generation (6G) millimeter wave (mmWave) communications. In RDARS-aided mmWave systems, the active and passive beamforming design and working mode configuration for reconfigurable elements are crucial for system performance. In this paper, we aim to maximize the weighted sum rate (WSR) in the RDARS-aided mmWave system. To take advantage of RDARS, we first design a reconfigurable codebook (RCB) in which the number and dimension of the codeword can be flexibly adjusted. Then, a low overhead beam training scheme based on hierarchical search is proposed. Accordingly, the active and passive beamforming for data transmission is designed to achieve the maximum WSR for both space-division multiple access (SDMA) and time-division multiple access (TDMA) schemes. For the TDMA scheme, the optimal number of RDARS transmit elements and the allocated power budget for WSR maximization are derived in closed form. Besides, the superiority of the RDARS is verified and the conditions under which RDARS outperforms RIS and DAS are given. For the SDMA scheme, we characterize the relationship between the number of RDARS connected elements and the user distribution, followed by the derivation of the optimal placement positions of the RDARS transmit elements. High-quality beamforming design solutions are derived to minimize the inter-user interference (IUI) at the base station and RDARS side respectively, which nearly leads to the maximal WSR. Finally, simulation results confirm our theoretical findings and the superiority of the proposed schemes.
\end{abstract}


%
\IEEEpeerreviewmaketitle

\section{Introduction}
With the ever-increasing requirements for communication performance, the sixth-generation (6G) is anticipated to meet the demands of ultra-high data rate transmission, ultra-low latency, as well as ultra-reliable connection. Millimeter wave (mmWave) communication has been regarded as a promising technology for 6G due to the huge available bandwidth \cite{xue_Survey}. 
However, mmWave signals suffer from severe attenuation and diffraction, which may limit its practical application\cite{NadezhdaChukhno_Survey}. 
To overcome the issue of weak penetration, 
a distributed antenna system (DAS) with multiple transmission and reception points connected to a single processing unit has been proposed to reduce outage and improve coverage \cite{Robert_DAS, Antonino_C_RAN}. 
In DAS, multiple antennas are geographically distributed throughout the network to provide the distributed gain, thereby achieving a high transmission rate. 
Various DAS architectures have been studied in different systems, such as cloud radio access networks, ultra-dense networks, and distributed cell-free multi-input and multi-output (MIMO) systems \cite{Antonino_C_RAN, Smruti_C_RAN, haiquan_survey, Andr_CF}.
However, DAS requires a large number of antennas and radio frequency (RF) chains, thus leading to high hardware cost and power consumption, especially for massive MIMO systems. 
To address these issues, reconfigurable intelligent surface (RIS) has been proposed as a low-cost technique in \cite{Ertugrul_RIS, xue_DoubleRIS, haiquanlu_RIS, AhmedElzanaty_RIS}, which can reshape the wireless propagation environment dynamically by adjusting the phase shift of each passive element. Owing to their appealing features, extensive studies have been devoted to this research direction for coverage enhancement, improving the performance of edge user equipments (UEs), and improving the sensing performance \cite{Shuhaozeng_RIS, haiquanlu_RIS, AhmedElzanaty_RIS}. It is also worth mentioning that RIS-aided system performance is limited by the multiplicative fading effect \cite{RuizheLong_Multifading}, and various RIS variants have been proposed, such as active RIS and hybrid RIS \cite{Peng_Power_constraint, Rafaela_HybridRIS, peng_semiRIS}, where the amplifiers are used to enhance the signal strength so as to benefit from the active beamforming gain. 
Similarly, to reap the high beamforming gain, a large number of elements are usually required, which incurs the prohibitively high overhead of the control link, rendering it difficult to meet the extremely high-performance requirements in 6G. 

To leverage the benefits and mitigate the limitations of RIS and DAS, a novel architecture termed reconfigurable distributed antenna and reflecting surface (RDARS) was proposed in \cite{ChengzhiMa_ANewArchi}, recently. RDARS has signal processing capabilities while reducing hardware costs and energy consumption, thanks to the cooperation among elements working in different modes, which improves the system's capacity and reliability \cite{zhang_RDARS, Wang_RDARS, ChengzhiMa_RDARS, jintao_ISAC_RDARS}. 
To be specific, RDARS consists of many reconfigurable elements, where each element is dynamically switched between two working modes: \textit{reflection mode} and \textit{connection mode}. In particular, for the reflection mode, RDARS element reduces to the conventional RIS passive element, while for the connection mode, RDARS serves as the distributed antennas connected to the base station (BS) via dedicated wires or fibers \cite{Robert_DAS}. 
By utilizing low-cost passive elements and existing DAS infrastructure, RDARS can reduce the hardware cost and provide satisfactory performance by introducing the appealing passive beamforming gain into the wireless communication system.
Furthermore, by dynamically switching working modes, RDARS enables more degree of freedom (DoF) and flexibility for system design.
Specifically, the theoretical performance analysis and experimental demonstrations have been carried out in \cite{ChengzhiMa_ANewArchi}.
In \cite{Wang_RDARS}, the mean-square-error has been minimized in the uplink MIMO communication system to improve the system transmission reliability. 
The radar output signal-to-noise ratio (SNR) has been maximized in the multi-user integrated sensing and communication (ISAC) system in \cite{zhang_RDARS}. 
Besides, two-timescale transceiver design has been investigated in a massive
MIMO uplink communication system in \cite{ChengzhiMa_RDARS}.

To fully reap the potential gains brought by the RDARS, it is crucial to properly design the reflection coefficients, operating mode configuration, and active transmit/receive beamforming.
However, such designs rely on the accurate channel state information (CSI).
Since RDARS consists of a large number of reflecting elements, thus incurring the high channel estimation overhead. 
In mmWave wireless systems, a practical approach for channel estimation is beam training (BT), such as codebook-based BT method \cite{xue_Survey, HuanHuang_TT, SungGeunYoon_Codebook, jiancheng_Codebook, DeyouZhang_CB}.
One straightforward method is to exhaustively search for all codewords in the codebook, but it faces extremely high training overhead. To address this issue, hierarchical search based on hierarchical codebooks has been proposed, which includes sector level phase and beam refinement phase \cite{JingheWang_Shutdown, ZhenyuXiao_HS, ChenhaoQi_HS, YuLu_HS, ChangshengYou_Static}. 
In \cite{ZhenyuXiao_HS}, an efficient hierarchical codebook by jointly exploiting sub-array and deactivation (turning-off) antenna processing techniques has been proposed. The authors of \cite{ChangshengYou_Static} proposed an efficient beam bin allocation and multi-beam training method to reach the comparable passive beamforming with a significantly low overhead. However, these approaches suffer from the effect of weak signal strength due to deactivating part of elements for BT. To tackle this issue, a hierarchical codebook has been proposed in \cite{ChenhaoQi_HS}, where all antennas were enabled to increase the success rate of beam alignment with the improved channel gain. The above codebooks are designed for conventional uniform linear arrays (ULAs) or uniform planar arrays (UPAs), where the spacing between two adjacent array elements is fixed.
In addition, the multiplicative channel in RIS-aided systems renders codebook-based BT more challenging due to passive architecture. In \cite{PeilanWang_RIS_Passive, ChenchengZhang_RIS_single_UE, Yuhao_Chen_RIS_single_UE}, the joint BT among BS, RIS, and UE is performed to obtain the cascaded channel. To this end, the three-stage BT method was proposed in a multi-RIS-aided system to conduct the empirical BT in \cite{HuanHuang_TT}. 
Moreover, with the CSI obtained by BT, codebook-based beamforming design has been extensively studied to reduce the complexity of beamforming design algorithm in \cite{HuanHuang_TT, Xing_Jia_Codebook_BF_MISO, Xu_Shi_Codebook_BF_single_UE, Baishuo_Lin_C_BF_MU}. 
However, existing codebook-based beamforming methods only consider single-user scenario, and each UE only is equipped with one single antenna in \cite{ HuanHuang_TT, Xing_Jia_Codebook_BF_MISO, Xu_Shi_Codebook_BF_single_UE, Baishuo_Lin_C_BF_MU}. 

Based on the above discussions, it is natural to investigate the potential performance gain of integrating RDARS into mmWave multi-user MIMO systems.
Regarding a RDARS-aided multi-user MIMO system, several fundamental issues remain unsolved. 
First, how to design a novel codebook to unleash the potential of RDARS in such systems? 
Since flexible mode switching brings an extra DoF, it is generally believed that exploiting appropriate mode configurations has a beneficial effect on system performance. Additionally, existing codebooks severely limit the performance of BT and data transmission due to the fixed phase difference between any adjacent antennas/elements for codewords.  
Taking the above factors into consideration, it is urgent to design a suitable codebook with reconfigurable codewords for RDARS.
Second, how to design a low overhead BT scheme in such systems? This question is driven by the fact that the overhead of BT is dominated by the cascaded channel estimation, where a large number of passive elements exaggerates this challenge.
Considering the dynamic working mode switching of RDARS elements, a flexible mode switching scheme may bring favorable effects on reducing the overhead.
Third, how to determine the number and placement positions of elements working in different modes to achieve the high weighted sum rate (WSR) with different user distributions for data transmission? A reconfigurable codebook is required due to the time-varying user distribution. Based on the user distribution obtained via the BT scheme, mode switching and the number and dimension of codewords can be determined. As such, a more flexible codebook-based beamforming scheme is desired to fully exploit the performance gain of RDARS for data transmission.

Motivated by the above issues, we focus on a RDARS-aided multi-user MIMO system by considering different multiple access schemes, where a RDARS is deployed to assist in the BT and downlink data transmission between multiple UEs and a BS.
Compared to the conventional RIS-aided system and DAS, the RDARS mode switching provides a new DoF to overcome the severe multiplicative fading and enhance the system capacity. 
On the other hand, the optimization variables, including transmit power budget, beamforming vectors, and mode switching matrix, are coupled in the newly added binary mode switching constraints, which renders the joint design of the beamforming and resource allocation for RDARS-aided system more challenging.
The main contributions are summarized as follows:
\begin{itemize}
 \item Firstly, we design a novel reconfigurable codebook (RCB) based on the discrete Fourier transform (DFT) codebook, where each codeword is related to the number and positions of antennas/elements, so as to cater to the dynamic configuration of elements at the RDARS. Specifically, the RCB is a two-dimensional (2D) architecture, which can degrade to a general beamsteering codebook when the inter-distance between elements is uniform. In addition, the number and dimension of the codeword in RCB can be dynamically adjusted according to the user distribution.
 \item Secondly, we propose a RCB-based joint low overhead BT and beamforming design algorithm. For the low overhead BT scheme, the hierarchical 2D RCB is generated to realize the efficient BT. Besides, the joint BT is separated into two parts among active antennas or elements by leveraging the active feature of connected elements and flexible RDARS element configurations.
 \item Lastly, the RCB-based active and passive beamforming designs are respectively considered in time-division multiple access (TDMA) and space-division multiple access (SDMA) schemes. For the TDMA scheme, we demonstrate the superiority of the RDARS-aided system in terms of WSR. Moreover, we derive the optimal power budget allocation and RDARS element allocation in closed form. Then, the impact of the system parameters on the WSR is characterized, e.g., path loss and the total number of RDARS elements. For the SDMA scheme, an alternative codeword is introduced to diminish inter-user interference (IUI) at the BS side. Then, the number and placement positions of RDARS transmit elements are optimized to mitigate IUI at the RDARS side. The IUI mitigation at both the BS and RDARS sides leads to nearly maximum WSR. Furthermore, we characterize the relationship between the number of RDARS transmit elements and angle-of-departures (AoDs) among the RDARS and users.
Simulation results verify the theoretical analysis and illustrate the effectiveness of the proposed schemes for the RDARS-aided systems. 
\end{itemize} 

The remainder of this paper is organized as follows. Section \ref{sec: system model} introduces the system model and problem formulation.  
Section \ref{sec: codebook design} presents a RCB design.
Section \ref{sec: low overhead BT Scheme and BF} provides a low overhead BT scheme based on the proposed RCB and proposes the codebook-based beamforming design for the SDMA and TDMA schemes. 
Simulation results are shown in Section \ref{sec: simulation}. Finally, Section \ref{sec: conclusion} concludes this paper.

\textit{Notations:} 
${\mathbb C}^ {d_1\times d_2}$ stands for the space of $d_1\times d_2$ complex-valued matrices. 
For a complex-valued scalar $x$, the modulus is denoted by ${| x |}$ and $\arg(x)$ denotes its phase.
For a real-valued scalar $x$, $\lfloor x \rfloor$ denotes the floor of $x$.
For a complex-valued vector $\bf x$, ${\left\| {\bf x} \right\|}$ represents the Euclidean norm of $\bf x$, ${\operatorname{diag}}(\bf x) $ denotes a diagonal matrix where the main diagonal elements are from $\bf x$, $[\mathbf{x}]_n$ denotes the $n$-th entry of $\mathbf{x}$. $\otimes$ and $\odot$ denote the Kronecker and Hadamard operators, respectively. For a matrix $\bf X$, ${\operatorname{tr}}( {\bf{X}} )$ and $\mathbf{X}_{[i,j]}$ respectively stand for its trace and the element in the $i$-th row and the $j$-th column, respectively. 
\section{System Model And Problem Formulation}\label{sec: system model}
\subsection{RDARS Setup}
A RDARS is composed of $N $ elements, denoted by the set, $\mathcal{N} = \{1, \cdots, N\}$, where the working mode of each element can be switched by the controller between the \textit{connection mode} and \textit{reflection mode}. 
Let a diagonal matrix $\mathbf{A}  \in \mathbb{Z}^{N\times N} $ denote the mode switching matrix, with the diagonal value of $\mathbf{A}$, i.e., $\mathbf{A}_{[i, i]}$, being 1 or 0, where $\mathbf{A}_{[i, i]}=1$ indicates that the $i$-th element works in connection mode as a transmit element (TE) or receive element (RE), and $\mathbf{A}_{[i, i]}=0$ represents the $i$-th element works in reflection mode as a passive element (PE). 
Moreover, the number of elements working in the connection and reflection modes are denoted by $a$ and $b$, respectively.
\subsection{System Model}
\begin{figure}[t] 
 \centering
 \includegraphics[width=0.35\textwidth]{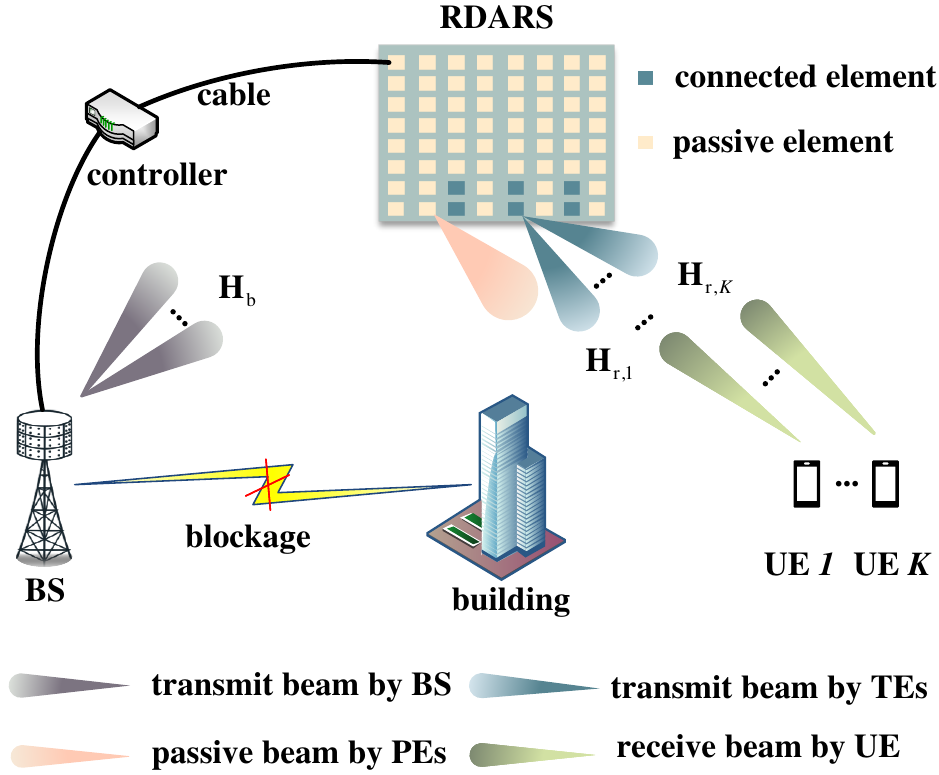}
  \vspace{-5pt}
 \caption{A RDARS-aided downlink multi-user MIMO architecture.}
 \label{fig: system architecture}
  \vspace{-15pt}
 \end{figure}

As shown in Fig. \ref{fig: system architecture}, we consider a RDARS-aided downlink communication system with one BS and $K$ UEs, where the BS is equipped with a uniform linear array (ULA) with $N_{\rm{t}}$ antennas and each UE is equipped with a ULA with $N_{\rm{u}}$ antennas. The RDARS is of uniform planar array (UPA) architecture with $N$ elements.
The beamforming matrix of the whole system is denoted by $\mathbf{F}_{\rm{R}} = \left[\mathbf{P}^{\rm{T}} _{\mathrm{B}}\mathbf{W}^{\rm{T}} , \mathbf{P}^{\rm{T}} _{\mathrm{R}}\mathbf{F}^{\rm{T}}\right]^{\rm{T}} \in \mathbb{C}^{ (N_{\mathrm{t}}+a)\times K}$, where $\mathbf{P}_{\mathrm{B}}= \textrm{diag}([\rho_{1}, \rho_{2}, \cdots, \rho_{K} ]) $, $\mathbf{P}_{\mathrm{R}} = \textrm{diag}([\widetilde{\rho}_{1}, \widetilde{\rho}_{2}, \cdots, \widetilde{\rho}_{K}])$, $\mathbf{W}=\left[\mathbf{w}_{1},\mathbf{w}_{2}, \cdots,\mathbf{w}_{K}\right]$ $\in \mathbb{C}^{N_{\rm{t}} \times K}$, and $\mathbf{F} = \left[ \mathbf{f}_{1}, \mathbf{f}_{2},\cdots, \mathbf{f}_{K}\right] \in \mathbb{C}^{a \times K}$ denote the transmit power budget allocation matrix at the BS, the transmit power budget allocation matrix at RDARS transmit elements, the beamforming matrix for BS, and the beamforming matrix for RDARS transmit elements, respectively. Meanwhile, $P_{\mathrm{B},k}$ and $P_{\mathrm{R},k} $ denote the transmit power budget for the $k$-th UE at each antenna of BS and the transmit power budget allocated to the $k$-th UE at each RDARS transmit element, with $\rho_{k} = \sqrt{P_{\mathrm{B},k}}$ and $\widetilde{\rho}_{k} = \sqrt{P_{\mathrm{R},k}}$, respectively.
Let ${\bf{s}} \in \mathbb{C}^{K \times 1}$ denote the transmit symbol, where $s_{k}, \forall k \in \mathcal{K} = \{1, \cdots, K\}$, satisfies $\mathbb{E}\{ {s}_{k}s_{k} ^{\mathrm{H}} \} = 1$ and $\mathbb{E}\{ s_{k} s_{i}^{\mathrm{H}} \} = 0, k \ne i$. The combined transmitted signal from the BS and RDARS is given by
\begin{equation} \label{equ: transmit signal for BS}
\begin{aligned}
 \mathbf{x} &= 
 \begin{bmatrix}
\mathbf{x}_{\mathrm{B}} \\
\mathbf{x}_{\mathrm{R}}
\end{bmatrix} =
 \begin{bmatrix}
\mathbf{W}\mathbf{P}_{\mathrm{B}} \\
\mathbf{F} \mathbf{P}_{\mathrm{R}}
\end{bmatrix}
\mathbf{s} = \mathbf{F}_{\mathrm{R}} \mathbf{s},
\end{aligned}
\end{equation}
where $\mathbf{x}_{\mathrm{B}}$ and $\mathbf{x}_{\mathrm{R}}$ denote the transmit signals from BS and from RDARS transmit elements, respectively. 
The channels spanning from the BS to RDARS and from the RDARS to the $k$-th UE are denoted by $\mathbf{H}_{\mathrm{\rm{\mathrm{b}}} } \in \mathbb{C}^{N \times N_{\rm{t}}}$ and $\mathbf{H}_{\mathrm{r},k} \in \mathbb{C}^{N_{\rm{u}}\times N}$, respectively. 
Let $\phi$ and $d$ denote the constant phase difference and the spacing between adjacent antennas/elements, respectively. The steering vector function is given by
$ \mathbf{a}(N,\phi) = [1, e^{\jmath\frac{2\pi}{\lambda} d \phi}, \cdots, e^{\jmath\frac{2\pi}{\lambda} d(N-1)\phi}]^{\rm{T}}$,
where $N$ is the array size, $\lambda$ represents the wavelength, $\jmath$ is the imaginary unit. 
For mmWave channels, 
the non-line-of-sight (NLoS) component is generally much weaker as compared to the dominant LoS component due to severe path loss, which thus can be ignored \cite{NLOS_Ignore}, \cite{Xianghao_NLOS}. Due to the high directivity of mmWave signals, the direct links between the BS and UEs are assumed to be severely blocked by obstacles. Considering the extended Saleh-Valenzuela model \cite{ChenhaoQi_HS}, let ${\theta}^{\rm{\mathrm{b}}_D}$, $\vartheta^{\rm{b_A}}$, and $\theta^{\rm{b_A}}$ denote the AoD, the azimuth and elevation angles-of-arrival (AoAs) from the BS to RDARS, respectively. In addition, let $\theta^{\mathrm{r_A}}_k$,  $\vartheta^{\mathrm{r_D}}_k$ and $\theta^{\mathrm{r_D}}_k$ denote the AoA, the azimuth and elevation AoDs from RDARS to the $k$-th UE, respectively. Therefore, the channels from the BS to RDARS and from RDARS to UE $k$ are expressed as 
\begin{equation} \label{equ: Channel BR}
 {\mathbf{H}}_{\mathrm{\mathrm{b}}} = \kappa_{{\rm{\mathrm{b}}}}{\widetilde{{\bf{a}}}}(N, \chi_{\rm{r}},\psi_{\rm{r}}){\bf{a}}^{\mathrm{H}}({{N_{\rm{t}}}}, \bar{\upsilon}),
\end{equation}
\begin{equation}\label{equ: Channel RU}
 {\mathbf{H}}_{\mathrm{r},k} = \kappa_{\mathrm{r},k}{\bf{a}}({{N_{\rm{u}}}}, \upsilon^{{\mathrm{r}}}_{k}) {\widetilde{{\bf{a}}}^{\mathrm{H}}}(N, \widetilde{\upsilon}_{k},\upsilon_{k}),
\end{equation}
where $\chi_{\rm{r}} = \cos\theta^{\rm{b_A}}$, $\psi_{\rm{r}} = \sin\vartheta^{\rm{b_A}}\sin\theta^{\rm{b_A}}$,
$\bar{\upsilon} = \cos {\theta}^{\rm{\mathrm{r}}_D}$, $\widetilde{\upsilon}_{k} = \cos \theta^{\mathrm{r_D}}_k$ ,
$\upsilon_{k} = \sin \vartheta^{\mathrm{r_D}}_k \sin \theta^{\mathrm{r_D}}_k$, and 
$\upsilon^{\mathrm{r}}_{k} = \cos\theta^{\rm{r_A}}_{k}$
denote the vertical virtual AoA, horizontal virtual AoA, virtual AoD from the BS to RDARS, vertical virtual AoD, horizontal virtual AoD, and virtual AoA from RDARS transmit elements to the $k$-th UE, respectively.
Let $(\kappa_{{\rm{\mathrm{b}}}})^2$ and $(\kappa_{{\rm{\mathrm{r}}},k})^2$ denote the large-scale path loss of the BS-RDARS link and the RDARS-UE $k$ link, respectively.
We define the 2D reconfigurable steering function as $\widetilde{{\bf{a}}}(N, \widetilde{\upsilon}_{k},\upsilon_{k}) = {\bf{a}}(N_{\rm{z}}, \widetilde{\upsilon}_{k}) \otimes {\bf{a}}(N_{\rm{y}}, \upsilon_{k})$, where $N_{\rm{z}}$ and $N_{\rm{y}}$ denote the numbers of elements along the $z$-axis and $y$-axis, respectively, with $N = N_{\rm{z}} N_{\rm{y}}$.
Therefore, the receive signal for the $k$-th UE is $\mathbf{y}^{\mathrm{s}}_{k} = \mathbf{H}_{k} \mathbf{F}_{\mathrm{R}} \mathbf{s} + \mathbf{n}_{k}$, where $\mathbf{H}_{k} = [\mathbf{H}_{\mathrm{\rm{\mathrm{r}}},k} \mathbf{G}\mathbf{H}_{\mathrm{\rm{\mathrm{b}}}} \;\; \mathbf{H}_{\mathrm{r},k}\widetilde{\mathbf{A}} ] \in \mathbb{C}^{N_{\mathrm{u}} \times (N_{\mathrm{t}} + a)}$ and ${\bf{n}}_{k}$ denote the effective channel and noise vector for the $k$-th UE with ${\bf{n}}_{k} \sim {\cal C}{\cal N}\left( {{\bf{0}},\sigma^2{{\bf{I}}_{N_{\rm{u}}}}} \right)$. We denote $\mathbf{A} =\widetilde{\mathbf{A}}\widetilde{\mathbf{A}}^{\mathrm{T}}$, where $\widetilde{\bf{A}} \in \mathbb{R}^{N \times a}$ includes all the columns of $\bf{A}$ that contain element 1. Let $\mathbf{G}$ denote the equivalent phase shift matrix with $\mathbf{G} = (\mathbf{I}_{N }-\mathbf{A} )\mathbf{\Phi}$, where $\mathbf{\Phi} = \operatorname{diag}(\bf v)$ with $\mathbf{v} = [ {e^{\jmath{\bar{\theta}_1}}}, {e^{\jmath{\bar{\theta}_2}}}, \cdots, {e^{\jmath{\bar{\theta}_N}}}]^{\mathrm{T}}$. Let ${\bar{\theta}_i}$ denote the phase shift of the $i$-th RDARS element. With the receive beamforming $\mathbf{u}_k$ at UE $k$, the signal-to-noise-plus-interference ratio (SINR) for the $k$-th UE is given by $\gamma_{k} = \frac{|\mathbf{u}^{\mathrm{H}}_k \mathbf{H}_{k} \mathbf{f}_{\mathrm{R},k}|^2 }{\sum_{ i \neq k}^{K}|\mathbf{u}^{\mathrm{H}}_k \mathbf{H}_{k} \mathbf{f}_{\mathrm{R},i }|^2 + \sigma^2 }$. Thus, the achievable data rate (bps/Hz) of the $k$-th UE is $R_{k} = \log_2 \left(1+ \gamma_k\right)$.

\subsection{Problem Formulation}
We aim to maximize the WSR by jointly optimizing the transmit beamforming matrix ${\bf{W}}$, power budget matrix ${\bf{P}}_{\rm{B}}$ at the BS, receive beamforming matrix $\mathbf{U} = [\mathbf{u}_1,\mathbf{u}_2, \cdots, \mathbf{u}_K]$, transmit beamforming matrix ${\bf{F}}$, phase shift matrix ${\bm{\Phi}}$, mode switching matrix ${\bf{A}}$ and power budget matrix ${\bf{P}}_{\rm{R}}$ at the RDARS. The problem can be formulated as  
\begin{subequations}\label{pro: sum rate}
 \begin{align}
 \mathop {\max }\limits_{\substack{\mathbf{W}, \mathbf{F}, \mathbf{A},\mathbf{\Phi},\\ \mathbf{U}, \mathbf{P}_{\mathrm{B}}, \mathbf{P}_{\mathrm{R}} }} \;
 & \;{\sum\nolimits_{k = 1}^K{w_{k}R_{k}}} 
 \\
 \textrm{s.t.}\;\;\;\;\;
 & N_{\rm{t}}\operatorname{tr}(\mathbf{P}_{\mathrm{B}}\mathbf{P}_{\mathrm{B}}^{\mathrm{H}}) + a\operatorname{tr}(\mathbf{P}_{\mathrm{R}}\mathbf{P}_{\mathrm{R}}^{\mathrm{H}}) \le P_{\rm{tot}}, \label{con: P}\\
 & \left\|\mathbf{w}_k\right\|^2 = 1,  \forall k\in \mathcal{K},\label{con: w}\\
 & \left\|\mathbf{f}_k\right\|^2 = 1,  \forall k\in \mathcal{K},\label{con: f}\\
 & \left|\mathbf{\Phi}_{[i,i]}\right|^2 = 1,  \forall i \in \mathcal{N},\label{con: Phi} \\
 & \mathbf{A}_{[i,i]} \in \{0,1\}, \forall i \in \mathcal{N}, \label{con: A}\\
 & \left\|\mathbf{u}_k\right\|^2 = 1,\; \forall k\in \mathcal{K},\label{con: u}
 \end{align}
\end{subequations}
where $w_{k}$ represents the weighted factor with respect to UE $k$ and ${P}_{\rm{tot}}$ denotes the total transmit power budget. 
It is observed that problem \eqref{pro: sum rate} is highly non-convex since the optimization variables are intricately coupled in the objective function and constraints. Moreover, the unit-modulus phase shift constraint \eqref{con: Phi} and working mode switching constraint \eqref{con: A} exacerbate the difficulties, yielding a mixed-integer nonlinear programming (MINLP) problem.
Furthermore, an efficient optimization algorithm deeply relies on the explicit CSI, which is difficult to obtain due to the large number of elements and antennas.
To tackle this problem, we first design a novel RCB. Then, a low overhead BT is proposed, followed by the high-quality solutions to the optimization problems in different MA schemes by exploiting inherent properties of RDARS.

\section{Reconfigurable Codebook Design}\label{sec: codebook design}
\begin{figure}[t] 
 \centering
 \includegraphics[width=0.35\textwidth]{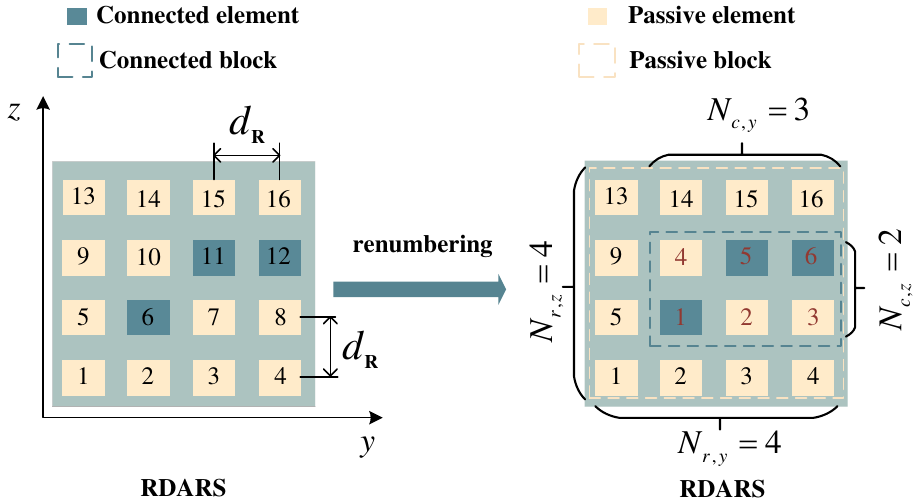}
 \vspace{-10pt}
 \caption{Illustration of the passive block and connected block.}
 \label{fig: RCB}
 \vspace{-15pt}
\end{figure}

To obtain the CSI by BT and perform the beamforming design for data transmission, the proper codebooks are essential for the BS, RDARS connected elements, passive elements, and UEs. 
However, considering the different working modes for RDARS elements, the conventional fixed codebook cannot be directly applicable to RDARS due to the non-uniform distance between any two elements working in the same mode after the flexible mode configuration. To this end, we propose a RCB with the non-uniform phase offset in each codeword and then provide a theoretical performance analysis. Besides, an example of the RCB is presented for clarity. 
Firstly, 
we consider a Cartesian coordinate system with the $x$-axis perpendicular to the RDARS plane, where the lower left corner is set as the reference point, with the inter-element center spacing given by $d_{\mathrm{R}}$. Based on the element coordinate, the reconfigurable steering function is given by
\begin{equation}
 \widehat{\mathbf{a}}(\mathsf{z}, N, \phi) =\left[e^{\jmath\frac{2\pi}{\lambda}z_1\phi},e^{\jmath\frac{2\pi}{\lambda}z_2\phi}, \cdots, e^{\jmath\frac{2\pi}{\lambda}z_N\phi}\right]^{\mathrm{T}},
 \label{equ: dynamic steering vector}
\end{equation}
where $\mathsf{z} = \left[z_1, z_2, \cdots, z_N\right]^{\mathrm{T}}$ denotes the coordinate vector of elements. 
A passive block or connected block is defined as the block surrounded by the outermost passive elements or connected elements, as shown in Fig. \ref{fig: RCB}.
Let $N_{\rm{r,z}}$ and $N_{\rm{r,y}}$ denote the total number of elements in the passive block along the $z$ and $y$ directions respectively. The total number of elements in a passive block and a connected block are $N_{\rm{r}} = N_{\rm{r,z}}N_{\rm{r,y}}$ and $N_{\rm{c}}=N_{\rm{c, z}}N_{\rm{c, y}}$, respectively.

For passive elements at the RDARS, the 2D RCB is designed as $\mathcal{C}_{\mathrm{R},2D} = \{\mathbf{r}_1,\cdots, \mathbf{r}_{\varrho_{\rm{r,z}}\times \varrho_{\rm{r,y}}}\}$, where $\mathbf{r}_{i} \in \mathbb{C}^{N_{\rm{r}} \times 1}$, $\varrho_{\rm{r,z}}$ and $\varrho_{\rm{r,y}}$ represent the $i$-th codeword, the resolution along the $z$-axis and that along the $y$-axis, respectively. Moreover, we denote $\mathcal{C}_{\rm{R,z}} = \{\mathbf{r}_{\mathrm{z},1}, \cdots, \mathbf{r}_{\mathrm{z}, \varrho_{\rm{r,z}}} \}$ and $\mathcal{C}_{\rm{R,y}} = \{\mathbf{r}_{\mathrm{y},1}, \cdots, \mathbf{r}_{y,\varrho_{\rm{r,y}}} \}$ as the one-dimensional (1D) passive RCB along the $z$ and $y$-axis, respectively. Note that the RDARS passive element is set to be a single beam in a given time slot, the orthogonality between any of the codewords in $\mathcal{C}_{\rm{R,z}}$ or $\mathcal{C}_{\rm{R,y}}$ is not required. This implies that a greater number of physical angles can be chosen as the codebook resolution increases. By applying the oversampling DFT codeword with $\mathbf{r}_{\mathrm{z},i} = \widehat{\mathbf{a}}(\mathsf{z}_{\rm{r}},N_{\rm{r, z}}, -1+\frac{2i-1}{\varrho_{\rm{r,z}}})$ and $\mathbf{r}_{y,i} = \widehat{\mathbf{a}}(\mathsf{y}_{\rm{r}},N_{\rm{r, y}},-1+\frac{2i-1}{\varrho_{\rm{r,z}}})$, we have $\varrho_{\rm{r,z}} \geq N_{\rm{r, z}}$ and $\varrho_{\rm{r,y}} \geq N_{\rm{r, y}}$.
Let $\bar{N}_{r,z}$ and $\bar{N}_{r,y}$ denote the numbers of the passive elements in the passive block along the $z$ and $y$ directions, respectively. The coordinates of RDARS passive elements on the $z$ axis and the $y$ axis are denoted by 
$\mathsf{z}_{\rm{r}} = \left[z_{\mathrm{r},1}, z_{\mathrm{r},2}, \cdots, z_{\mathrm{r},N_{\rm{r, z}}}\right]^{\mathrm{T}}$ and $\mathsf{y}_{\rm{R}} = \left[y_{\mathrm{r},1}, y_{\mathrm{r},2}, \cdots, y_{\mathrm{r},N_{\rm{r, y}}}\right]^{\mathrm{T}}$, respectively. 
The set of the element positions in the passive block is denoted by $\mathcal{P}_{\rm{R}} = \{(x_0, y_{\mathrm{r},m}, z_{\mathrm{r},n})| m = 1,\cdots, N_{\rm{r, y}}, n=1,\cdots, N_{\rm{r, z}} \}$, where $x_0$ is the coordinate of the entire RDARS plane in the $x$ direction. Therefore, we have $\mathbf{r}_i = \operatorname{vec}(\bar{\mathbf{A}}^{\mathrm{r}}) \odot (\mathbf{r}_{z,\widetilde{i}} \otimes \mathbf{r}_{y,\bar{i}})$, where ${\bar{\mathbf{A}}^{\mathrm{r}}} \in \mathbb{Z}^{N_{\rm{r}} \times N_{\rm{r}}}$ is a diagonal matrix, with ${{\bar{\mathbf{A}}}^{\mathrm{r}}}_{[i,i]} \in \{0, 1\}$,  $i = \tilde{i}\bar{i}$ and $\tilde{i} \geq \bar{i}$. After renumbering the elements in the passive block, the corresponding element operating in connection mode is set to 0. For example, in Fig. \ref{fig: RCB}, ${\bar{\mathbf{A}}^{\mathrm{r}}}_{[6,6]}$, ${\bar{\mathbf{A}}^{\mathrm{r}}}_{[11, 11]}$ and ${\bar{\mathbf{A}}^{\mathrm{r}}}_{[12, 12]}$ are set to 0.

On the other hand, connected elements at the RDARS in practice form a distributed antenna array. Similar to RDARS passive elements, the 2D RCB for connected elements is denoted by $\mathcal{C}^{\mathrm{c}}_{\mathrm{R},2D} = \{\mathbf{c}_{1},\cdots, \mathbf{c}_{\varrho_{\rm{c,z}}\times \varrho_{\rm{c,y}}}\}$, where $\mathbf{c}_{i} \in \mathbb{C}^{N_{\rm{c}} \times 1}$, $\varrho_{\rm{c,z}}$ and $\varrho_{\rm{c,y}}$ represent the $i$-th codeword, the resolution along the $z$-axis and that along $y$-axis, respectively. 
Let $N_{\rm{c, z}}$ and $N_{\rm{c, y}}$ denote the total number of elements in the connected block along the $z$ and $y$ directions, respectively.
Specifically, we denote $\mathcal{C}^{\mathrm{c}}_{\rm{R,z}} = \{\mathbf{c}_{\mathrm{z},1}, \cdots, \mathbf{c}_{\mathrm{z},\varrho_{\rm{c,z}}} \}$ and $\mathcal{C}^{\mathrm{c}}_{\rm{R,y}} = \{\mathbf{c}_{\mathrm{y},1}, \cdots, \mathbf{c}_{\mathrm{y},\varrho_{\rm{c,y}}} \}$ as the 1D connected RCB along the $z$ and $y$-axis, respectively. Moreover, we have $\mathbf{c}_{{\rm{z}}, i} = \widehat{\mathbf{a}}(\mathsf{z}_{\rm{c}},N_{\rm{c,z}},-1+\frac{2i-1}{\varrho_{\rm{c,z}}})$ and $\mathbf{c}_{\mathrm{y}, i} = \widehat{\mathbf{a}}(\mathsf{y}_{\mathrm{c}},N_{\rm{c, y}},-1+\frac{2i-1}{\varrho_{\rm{c,z}}})$. 
The set of element positions in the connected block is denoted by $\mathcal{P}_{c} = \{(x_0, y_{\mathrm{c},m}, z_{\mathrm{c},n})| m = 1,\cdots, N_{\rm{c, y}}, n=1,\cdots, N_{\rm{c,z}} \}$. 
In addition, we have $\mathbf{c}_i =  \frac{1}{\sqrt{a}}\operatorname{vec}(\bar{\mathbf{A}}^{\mathrm{c}}) \odot (\mathbf{c}_{\mathrm{z},\widetilde{i}} \otimes \mathbf{c}_{y,\bar{i}})$, where $a$ is equal to the number of elements in $\operatorname{vec}(\bar{\mathbf{A}}^{\mathrm{c}})$ with value 1, and the definition of $\bar{\mathbf{A}}^{\mathrm{c}}$ is similarly to $\bar{\mathbf{A}}^{\mathrm{r}}$. 
In general, multiple beams should be generated by connected elements for the simultaneous signal transmission or reception. Therefore,
to guarantee the orthogonality of any two codewords, we should have 
\begin{align}
    \label{equ: normlized c}
    |\mathbf{c}^{\rm{H}}_{i} \mathbf{c}_{j}|
    & = {||\operatorname{vec}(\bar{\mathbf{A}}^{\rm{c}})||^2}|(\mathbf{c}^{\rm{H}}_{\mathrm{z},\widetilde{i}} \otimes  \mathbf{c}^{\rm{H}}_{{\rm{y}},\bar{i} }) (\mathbf{c}_{\mathrm{z},\widetilde{j}} \otimes  \mathbf{c}_{\mathrm{y},\bar{j} })| / ({a N_{\rm{c}}}) \nonumber \\
    & = |\mathbf{c}^{\rm{H}}_{\mathrm{z},\widetilde{i}} \mathbf{c}_{{\rm{z}},\widetilde{j}}|
    |\mathbf{c}^{\rm{H}}_{\mathrm{y},\bar{i}} \mathbf{c}_{{\rm{y}},\bar{j}}| / {N_{\rm{c}}}\nonumber\\
    & = |\sum^{a_{\rm{z}}}_{n=1}e^{\jmath\frac{2\pi}{\lambda}\frac{2(\widetilde{j}-\widetilde{i})}{\varrho_{\rm{c,z}}}\widetilde{z}_{c,n}}| |\sum^{a_{\rm{y}}}_{m=1}e^{\jmath\frac{2\pi}{\lambda}\frac{2(\bar{j}-\bar{i})}{\varrho_{\rm{c,y}}}\widetilde{y}_{\mathrm{c},m}}| / a\nonumber\\
    & = \delta_{ij},
\end{align}
where $\delta_{ij}=1$ when $i=j$ and $\delta_{ij}=0$ otherwise. The last equation holds if and only if $\varrho_{\rm{c,z}} = a_{\rm{z}}$ and $\varrho_{\rm{c,y}} = a_{y}$. This result shows that the resolution of the connected RCB is limited by the number of RDARS connected elements. 
Furthermore, the resolution of the connected RCB depends on $a_{\rm{z}}$ and $a_{y}$, while is independent of the distribution of connected elements. The reason is that each connected element is required to be individually connected with a dedicated RF chain, and the number of RF chains determines the capability of simultaneous beam transmission.
It is also worth mentioning that the RCB can be designed to achieve flexible working mode configurations, and its size can be dynamically adjusted according to the UE distribution.
In particular, the 2D RCB reduces to the conventional fixed codebook when $\mathsf{z} = \left[0,d_{\mathrm{R}},\cdots, \left(N_{\rm{z}} - 1\right)d_{\mathrm{R}} \right]$ and $\mathsf{y} = \left[0,d_{\mathrm{R}},\cdots, \left(N_{\rm{y}} - 1\right)d_{\mathrm{R}} \right]$. As a result, the RCB is of high flexibility by relaxing the element spacing $d_{\rm{R}}$ to an arbitrary value.

\section{RCB-based Low Overhead BT and Beamforming Design}\label{sec: low overhead BT Scheme and BF}
In this section, we propose a low overhead BT scheme to obtain the CSI, followed by a low-complexity active and passive beamforming design. 
\subsection{Low Overhead BT Scheme}
\begin{figure}[t]
 \centering
 \includegraphics[width=0.4\textwidth]{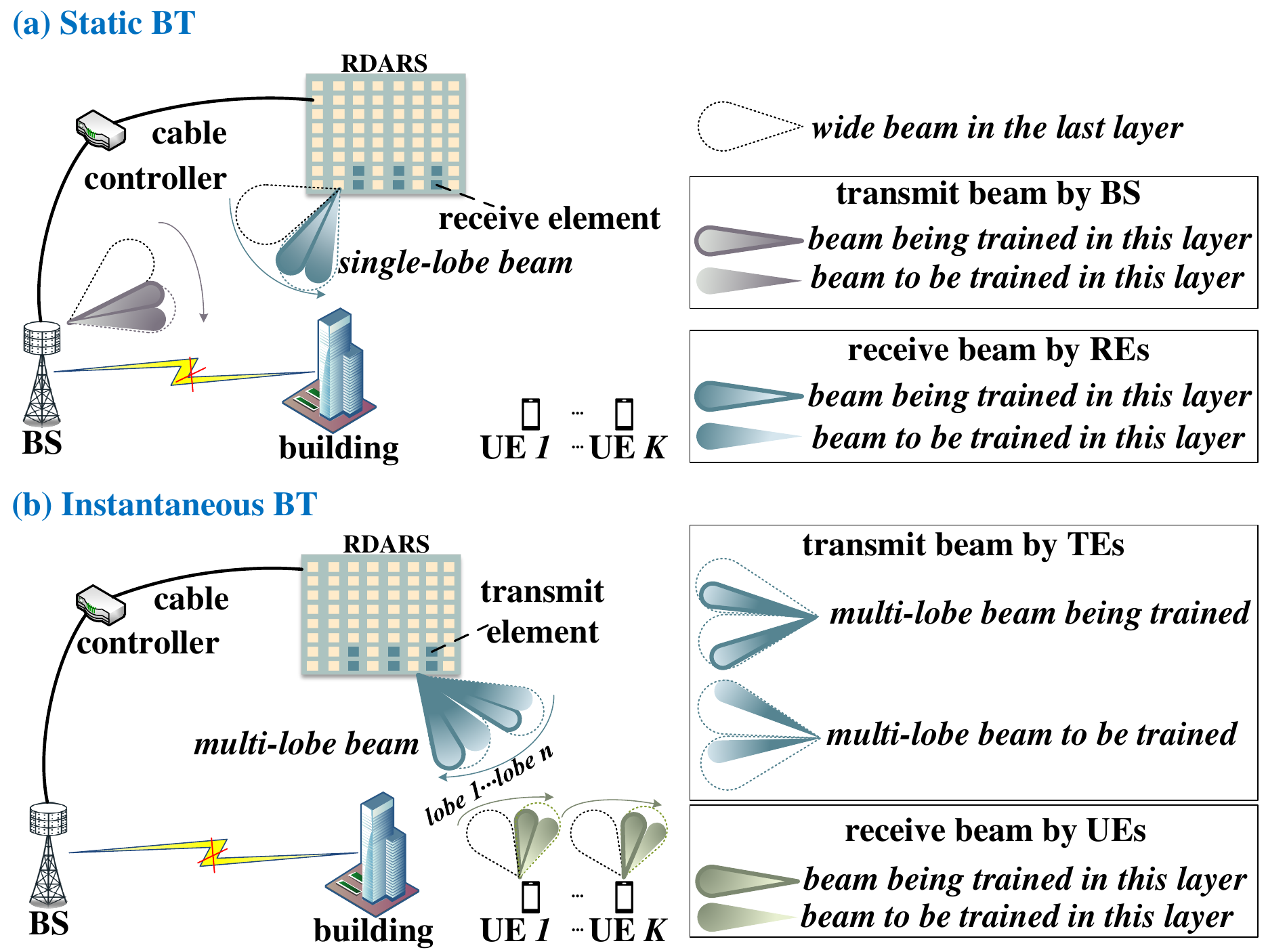}
 \caption{Illustration of the low overhead BT for RDARS-aided system.}
 \label{fig: HS BT}
 \vspace{-15pt}
\end{figure} 
In this subsection, we propose a low overhead BT scheme for the RDARS-aided system to obtain the required CSI, by applying the hierarchical RCB to the BS, UEs, and RDARS elements.

Let $\mathcal{C}_{\mathrm{B}}$ and $\mathcal{C}_{\mathrm{U}}$ denote the hierarchical codebooks for BS and UE, respectively, where $\mathcal{C}_{\mathrm{B}} = \{\mathbf{b}_{1},\mathbf{b}_2, \cdots, \mathbf{b}_{2N_{\rm{t}}-2} \} $ and $\mathcal{C}_{\mathrm{U}} = \{\mathbf{e}_{1},\mathbf{e}_2, \cdots, \mathbf{e}_{2N_{\rm{u}}-2} \}$. Let $\mathbf{b}_i$ and $\mathbf{e}_i$ denote the $i$-th codeword in the hierarchical codebook for BS and UE, respectively. To reduce the overhead, the simultaneous multi-user hierarchical BT among transmit elements and UEs is performed.

Before illustrating the BT procedure, the hierarchical codebooks applied for the BS, RDARS, and UEs are first designed, respectively. Specifically, for BS, UEs, and RDARS receive elements, the hierarchical codebook consists of $\log_M N_{\rm{t}}$ layers, where the $l$-th layer includes $2^l$ codewords. Each codeword covers a certain width in the angle domain, where the codeword in the upper layer with a wider beam coverage is divided into $M$ child codewords in the lower layer with a narrower beam coverage. By contrast, the hierarchical RCB for RDARS transmit elements is constructed based on the RCB, where the multi-lobe codeword in each layer is generated during the BT procedure \cite{ChenhaoQi_HS}. Thus, the codeword set at the bottom layer of the hierarchical RCB is $\mathcal{C}^{\mathrm{c}}_{\mathrm{R},2D}$. 

Next, we propose a low overhead BT approach. In practice, the BS-RDARS link is quasi-static since the locations of the BS and RDARS are fixed, and thus the BT between BS and RDARS only needs to be performed once during a large time scale \cite{HuanHuang_TT, ChangshengYou_Static}, termed ``\textit{Static BT}''. By contrast, the BT among RDARS connected elements and UEs is termed as ``\textit{Instantaneous BT}'', as shown in Fig. \ref{fig: HS BT}. 
Since the RDARS size is much smaller than the link distance, the AoA and AoD for the passive elements are approximated to be the same as those for connected elements. By connecting the independent RF chain, connected elements can process
signals. Therefore, the joint BT for BS, passive elements, transmit elements, and UEs is separated into two independent, as elaborated below.

As illustrated in Fig. \ref{fig: HS BT}, firstly, the RDARS elements working in the receive mode serve as the receive antennas, which are regarded as a part of the antenna array at the BS. The static BT is performed between the receive elements and BS based on the received signal strength. To be specific, the BS sequentially switches the transmit beams by layer and the receive elements search $M$ receive beams in each layer, thus determining the optimal transmit codeword $\widetilde{\mathbf{b}}^{*}_{k}$ and passive receive codeword $\widetilde{\mathbf{r}}^{*}_{\mathrm{r}, k}$.
Secondly, the elements working in the receive mode are switched to transmit elements. The transmit elements sweep beams sequentially when UEs receive the signal from transmit elements simultaneously, yielding the transmit beamforming vector for transmit elements $\widetilde{\mathbf{r}}^{*}_{t,k} = \widetilde{\mathbf{c}}^{*}_{k}$ and the receive beamforming vector for the $k$-th UE $\widetilde{\mathbf{u}}^{*}_{k} = \widetilde{\mathbf{e}}^{*}_{k}$. 
Based on the determined beams, the large-scale path loss $\kappa^2_{\mathrm{b}}$ and $\kappa^2_{\mathrm{r}, k}$ can be estimated according to the received signal strength.
Then, the channel can be reconstructed based on the obtained angles at the bottom of the codebooks and large-scale path loss, so as to use for the data transmission \cite{ChenhaoQi_HS}.

\subsection{RCB-Based Beamforming Design}\label{sec: max sumrate}
With the estimated CSI, the optimal beamforming vectors for data transmission need to be chosen from the codebooks.
Therefore, problem \eqref{pro: sum rate} is reformulated as
\begin{subequations}\label{pro: sum rate based on codebook}
 \begin{align}
 \mathop {\max }\limits_{\substack{\mathbf{W}, \mathbf{F}, \mathbf{A},\mathbf{\Phi},\mathbf{U}, \mathbf{P}_{\mathrm{B}}, \mathbf{P}_{\mathrm{R}} }}
 & \;\;{\sum\nolimits_{k = 1}^K {\omega_k{R_{k}} }}
 \\
 \;\;\textrm{s.t.}\;\;\;\;\;\;\;\;
 & \mathbf{w}_{k} \in \mathcal{C}_{\rm{B}},  \forall k\in \mathcal{K},\label{con: w CB}\\ 
 &\mathbf{f}_{k} \in \mathcal{C}^{\rm{c}}_{\rm{R,2D}},  \forall k\in \mathcal{K},\label{con: f CB}\\
 & \mathbf{u}_{k} \in \mathcal{C}_{\rm{U}}, \forall k\in \mathcal{K}, \label{con: u CB}\\
 & \mathbf{\Phi}= \textrm{diag}(\mathbf{r}), \mathbf{r} \in \mathcal{C}_{\rm{R,2D}},\label{con: Phi CB} \\
 & \eqref{con: P}, \eqref{con: A}.
 \end{align}
\end{subequations}
It is observed that problem \eqref{pro: sum rate based on codebook} can be solved by enumerating all possible beam pair combinations, however, this causes high computational complexity. Besides, the constraint \eqref{con: A} exacerbates the difficulty due to the flexible element configurations at the RDARS. 

To overcome these issues, we propose effective algorithms based on the RCB for data transmission, by considering the TDMA and SDMA schemes, respectively. 
\subsubsection{TDMA} \label{sec: TDMA}
For the TDMA-based scheme, 
problem \eqref{pro: sum rate based on codebook} can be rewritten in a more tractable form as
\begin{subequations}\label{pro: sum rate based on codebook TDMA}
 \begin{align}
 \mathop {\max }\limits_{\substack{\{\mathbf{w}_k\}, \{\mathbf{f}_k\}, \{\mathbf{A}_k\},\{\mathbf{\Phi}_k\},\\ \{\mathbf{u}_k\}, \{P_{\mathrm{B},k}\}, \{P_{\mathrm{R},k}\}}}
 & \;{\sum\nolimits_{k = 1}^K {w_k\widetilde{R}^{T}_k} }
 \\
  \;\textrm{s.t.}\;\;\;\;\;\;\;\;\;\;
  & \eqref{con: w CB}, \eqref{con: f CB}, \eqref{con: u CB},\\
 & N_t{\rho}^2_{k} + a\widetilde{\rho}^2_{k} = P_{\rm{tot}},\forall k \in \mathcal{K}, \label{con: power in TDMA}\\
   & \mathbf{A}_{k[i,i]} \in \{0,1\}, \forall k \in \mathcal{K}, \forall i \in \mathcal{N}, \label{con: A in TDMA}\\
    & \mathbf{\Phi}_k = \textrm{diag}(\mathbf{r}_k), \mathbf{r}_k \in \mathcal{C}_{\rm{R,2D}},\label{con: Phi CB in TDMA}
 \end{align}
\end{subequations}
where $\widetilde{R}^{T}_k=\log_2(1+|\rho_{k} \mathbf{u}^{\mathrm{H}}_{k}\mathbf{H}_{\mathrm{r},k}({\bf{I}}_{N} - {\bf{A}}_k)\mathbf{\Phi}_k \mathbf{H}_{\mathrm{\mathrm{b}}} \mathbf{w}_{k} + \widetilde{\rho}_{k}\mathbf{u}^{\mathrm{H}}_{ k}\mathbf{H}_{\mathrm{r},k} \widetilde{\mathbf{A}}\mathbf{f}_{k}|^2 / \sigma^2)$ denotes achievable rate of UE $k$. Let $\mathbf{A}_k$ and $\mathbf{\Phi}_k$ denote the mode switching matrix and phase shift matrix for the $k$-th UE, respectively.
In particular, for the special case of the same weighting factor and total transmit power budget for each UE, problem \eqref{pro: sum rate based on codebook TDMA} for the $k$-th UE is reduced to 
\begin{subequations}\label{pro: single user in TDMA}
    \begin{align}
    \mathop {\max }\limits_{\substack{\mathbf{w}_k, \mathbf{f}_k, \mathbf{A}_k,\mathbf{u}_k, \\ \mathbf{\Phi}_k, {\rho}_{k}, \widetilde{\rho}_{k}}}
    & \!|\rho_{k} \mathbf{u}^{\mathrm{H}}_{k}\mathbf{H}_{\mathrm{r},k}\widetilde{{\bf{G}}}_k\mathbf{\Phi}_k \mathbf{H}_{\mathrm{\mathrm{b}}} \mathbf{w}_{k}\!\! + \!\!\widetilde{\rho}_{k}\mathbf{u}^{\mathrm{H}}_{ k}\mathbf{H}_{\mathrm{r},k} \widetilde{\mathbf{A}}\mathbf{f}_{k}|^2 \label{OF}\\
    \textrm{s.t.}\;\;\;\;
    &\eqref{con: w CB}, \eqref{con: f CB}, \eqref{con: u CB},  \eqref{con: power in TDMA}, \eqref{con: A in TDMA}, \eqref{con: Phi CB in TDMA},
    \end{align}
\end{subequations}
where $\widetilde{{\bf{G}}}_k = {\bf{I}}_{N} - {\bf{A}}_k$.
It can be observed that problem \eqref{pro: single user in TDMA} is a non-convex optimization problem due to the coupled variables and the codeword selection constraints \eqref{con: w CB}, \eqref{con: f CB}, \eqref{con: u CB}, and \eqref{con: Phi CB in TDMA}.
A closer look at \eqref{pro: single user in TDMA} shows that the objective function satisfies
\begin{align}
&|\rho_{k}\mathbf{u}^{\mathrm{H}}_{ k}\mathbf{H}_{\mathrm{r},k}\widetilde{\bf{G}}_k\mathbf{\Phi} \mathbf{H}_{\mathrm{\mathrm{b}}} \mathbf{w}_{k} + \widetilde{\rho}_{k}\mathbf{u}^{\mathrm{H}}_{ k}\mathbf{H}_{\mathrm{r},k} \widetilde{\mathbf{A}}\mathbf{f}_{k}| \nonumber\\
& \leq \rho_{k}|\mathbf{u}^{\mathrm{H}}_{k}\mathbf{H}_{\mathrm{r},k}\widetilde{\bf{G}}_k\mathbf{\Phi} \mathbf{H}_{\mathrm{\mathrm{b}}} \mathbf{w}_{k} |+\widetilde{\rho}_{k}| \mathbf{u}^{\mathrm{H}}_{k}\mathbf{H}_{\mathrm{r},k} \widetilde{\mathbf{A}}\mathbf{f}_{k}|, \label{equ: two parts of OF}
\end{align}
where the equality holds if and only if the two items in \eqref{equ: two parts of OF} have the identical phase. By leveraging this property and maximum ratio transmission (MRT)/maximum ratio combining (MRC) beamforming, the closed-form beamforming expressions can be obtained, including $\{{\bf{w}}_k\}$, $\{\mathbf{f}_k, {\bf{u}}_k\}$, and $\{\mathbf{\Phi}_k\}$.
For given ${\bf{A}}_k$, $\rho_{k}$, and $\widetilde{\rho}_{k}$, problem \eqref{pro: single user in TDMA} becomes
\begin{subequations}\label{pro: single user fixed A, P_B, P_R}
    \begin{align}
    \mathop{\max}\limits_{\substack{\mathbf{w}_k, \mathbf{f}_k, \mathbf{u}_k, \mathbf{\Phi}_k}} & |\rho_{k} \mathbf{u}^{\mathrm{H}}_{k}\mathbf{H}_{\mathrm{r},k}{\bf{G}}\mathbf{\Phi}_k \mathbf{H}_{\mathrm{\mathrm{b}}} \mathbf{w}_{k} \!\!+ \!\!\widetilde{\rho}_{k}\mathbf{u}^{\mathrm{H}}_{ k}\mathbf{H}_{\mathrm{r},k} \widetilde{\mathbf{A}}\mathbf{f}_{k}|^2 \\
    \textrm{s.t.}\;\;\;\;\;
    & \eqref{con: w CB}, \eqref{con: f CB}, \eqref{con: u CB}, \eqref{con: Phi CB in TDMA}.
    \end{align}
\end{subequations}
Given $\{\mathbf{f}_k, {\bf{u}}_k\}$ and $\mathbf{\Phi}_k$, problem \eqref{pro: single user fixed A, P_B, P_R} is reduced to
\begin{equation}\label{pro: single user_w}
	\mathop {\max }\nolimits_{ \{{\bf{w}}_{k}\} }\left| {\bf{a}}^{\mathrm{H}}({N_{\rm{t}}}, \bar{\upsilon}) {\bf{w}}_{ k} \right|^2 
\;\ \textrm{s.t.}\;\eqref{con: w CB}.
\end{equation}
The optimal solution to problem \eqref{pro: single user_w} is given by $\mathbf{w}_{k}^{*}=\mathbf{b}_{k}^*=\mathop{\arg\min}\nolimits_{\mathbf{b}_{n}} ||\mathbf{b}_{n} - \widetilde{\mathbf{w}}_{k}^{*}||$, with  $\widetilde{\mathbf{w}}_{k}^{*} ={\bf{a}}({{N_{\rm{t}}}}, \bar{\upsilon}) /\sqrt{N_{t}}$. 
Given $\{\mathbf{w}_k\}$ and $\mathbf{\Phi}_k$, problem \eqref{pro: single user fixed A, P_B, P_R} is reduced to
\begin{equation}\label{pro: single user_f}
\mathop {\max }\limits_{ {\bf{f}}_{ k}, {\bf{u}}_{ k}}
|\mathbf{u}^{\mathrm{H}}_{ k}{\bf{a}}({N_{\rm{u}}}, \upsilon^{r}_{\mathrm{R},k}) \widetilde{\bf{a}}^{\mathrm{H}}({N}, \widetilde{\upsilon}_k,{\upsilon}_k)\widetilde{\mathbf{A}} {\bf{f}}_{k} |^2 
\textrm{s.t.}\eqref{con: f CB}, \eqref{con: u CB}.
\end{equation}
The optimal receive beam for the $k$-th UE is $\mathbf{u}_{k}^*=\mathbf{e}_{k}^* =\mathop{\arg\min}\nolimits_{\mathbf{e}_{n}}\|\mathbf{e}_{n} - \mathbf{a}({N_{\rm{u}}}, \upsilon^{r}_{\mathrm{R},k}) / \sqrt{N_{\rm{u}}}\|$. 
For given $\mathbf{A}_k$, the optimal transmit beam for each UE at the RDARS side is chosen from the 2D RCB $\mathcal{C}^{\mathrm{c}}_{\mathrm{R}}$. Besides, for the connected block, we denote the equivalent codeword $\widetilde{\mathbf{c}} = \bar{\mathbf{A}}^{\mathrm{ce}}\mathbf{c}$, where $(\bar{\mathbf{A}}^{\mathrm{ce}})^{\rm{T}} \in \mathbb{R}^{a\times N_c}$ consists of all the columns of $\bar{\mathbf{A}}^{c}$ that contain element 1. 
Then, the optimal transmit beamforming vector for the $k$-th UE is $ {{\bf{f}}_k} = \widetilde{\mathbf{c}}_{k}^* =\mathop{\arg\min}\nolimits_{\bar{\mathbf{A}}^{\mathrm{ce}}\mathbf{c}_{n}}\|\bar{\mathbf{A}}^{\mathrm{ce}}\mathbf{c}_{n} - \breve{\mathbf{a}}(\mathsf{z}_t\otimes\mathsf{y}_t, a,\widetilde{\upsilon}_k,{\upsilon}_k) / \sqrt{a}\|$, where  $\mathsf{z}_t$ and $\mathsf{y}_t$ respectively denote the coordinate of transmit elements along $z$ and $y$ axis, and $\breve{\mathbf{a}}(\mathsf{z}_t\otimes\mathsf{y}_t, a,\widetilde{\upsilon}_k,{\upsilon}_k) = \widehat{\bf{a}}(\mathsf{z}_t, a_{\rm{z}}, \widetilde{\upsilon}_k) \otimes \widehat{\bf{a}}(\mathsf{y}_t, a_{\rm{y}},{\upsilon}_k) $.
Therefore, the upper bound gain for RDARS transmit elements is ${a}$.
For the passive beamforming matrix, the subproblem w.r.t. ${\bf{\Phi}}_k$ is  
\begin{align}\label{pro: single user_Phi}
	 \mathop{\max}\nolimits_{ {\{\bf{\Phi}}_k\}} |\widetilde{{\bf{a}}}^{\mathrm{H}}({N}, \widetilde{\upsilon}_k,{\upsilon}_k )\widetilde{\bf{G}}\mathbf{\Phi}_k \widetilde{\mathbf{a}}({N}, \chi_{\rm{R}}, \psi_{\rm{R}}) |^2
          \;\; \textrm{s.t.}\; \eqref{con: Phi CB in TDMA}.
\end{align}  
To tackle this problem, we reconstruct the passive beamforming vector based on the optimal RDARS transmit beam $\widetilde{\mathbf{r}}^{*}_{t,k}$ and the optimal RDARS receive beam $\widetilde{\mathbf{r}}^{*}_{r,k}$ obtained during BT. 
Specifically, the optimal passive beamforming vector is given by $\widetilde{\mathbf{v}}_{k,n}^{*} =e^{\jmath([\arg(\widetilde{\mathbf{r}}_{\mathrm{t}, k}^*)]_{n}-[\arg(\widetilde{\mathbf{r}}_{\mathrm{r}, k}^*)]_n ) }$. Then, the optimal passive beam is $\mathbf{r}_{k}^* =\mathop{\arg\min}\nolimits_{\mathbf{r}_{n}}\|\mathbf{r}_{n} -\widetilde{\mathbf{v}}_{k}^{*} \|$. 

With the optimized beamforming vectors, the achievable rate w.r.t. ${\bf{A}}_k$ and $P_{\mathrm{B}, k}$, and $P_{\mathrm{R}, k}$ for the $k$-UE is given by
\begin{equation} \label{equ: max rate for single user}
R^{*}_{k}=\log_2( 1 + \frac{ \left({G}_{\mathrm{b},k} + {G}_{\mathrm{r},k}\right)^2}{\sigma^2}),
\end{equation}
where ${G}_{\mathrm{R},k}=\kappa_{\mathrm{r}, k}\sqrt{P_{\mathrm{R},k} N_{\rm{u}} a} $ and ${G}_{\mathrm{B},k} = \kappa_{\mathrm{br}, k}(N-a)\sqrt{{P}_{\mathrm{B},k}N_{\rm{u}}N_{\rm{t}}}$, with $ \kappa_{\mathrm{br},k} =  \kappa_{\mathrm{b}} \kappa_{\mathrm{r},k}$.
It is observed from \eqref{equ: max rate for single user} that the achievable rate is independent of the placement positions of RDARS transmit elements.
Since each UE occurs in different time slots for the TDMA scheme, the single-user case is directly extended to the multi-user case for the TDMA scheme. To this end, we first consider the single-user case and ignore the index $k$ for the sake of brevity.
Based on \eqref{equ: max rate for single user}, the achievable rate of the single UE is 
\begin{equation} \label{equ: rate for single UE ignore k}
{R}_{\textrm{s}}= \log_2 (1 + \frac{ (G_{\rm{b}} + G_{\rm{r}} )^2 }{\sigma^2 }),
\end{equation}
where ${G}_{\rm{R}}=\kappa_{\mathrm{r}}\sqrt{P_{\mathrm{R}} N_{\rm{u}} a} $ and ${G}_{\mathrm{B}} = \kappa_{\mathrm{br}}(N-a)\sqrt{{P}_{\mathrm{B}}N_{\rm{u}}N_{\rm{t}}}$.
Given ${\bf{w}}^*$, ${\bf{f}}^*$, ${\bf{u}}^*$ and ${\bf{\Phi}}^*$, problem \eqref{pro: single user in TDMA} is reduced to
\begin{align}\label{pro: optimize power and A in TDMA}
    \mathop{\max}\limits_{P_{\mathrm{R}}, P_{\mathrm{B}}, a} 
	  \; R_{s}\; \;\;\;
      \textrm{s.t.}\; N_tP_{\mathrm{B}} + aP_{\mathrm{R}}  = P_{\rm{tot}},
      0\leq a \leq N.
\end{align}
It is observed that problem \eqref{pro: optimize power and A in TDMA} is convex for given $a$, thus we have the following proposition.
\begin{Proposition}\label{propo: optimal power allocation} 
Based on the optimal beamforming vectors, 
the optimal power budget allocation is given by
\begin{equation} \label{equ: P_R P_B in Propo 1}
    {P^{*}_{\mathrm{R}}} =
\begin{cases}
    0, \; a=0, \\
    \frac{{{P_{\rm{tot}}}}}{{{\kappa^2_{\mathrm{b}}}{{( {N - a})}^2}a + a}}, \; 0<a \leq N,
\end{cases}
\end{equation}
and $P^{*}_{\rm{B}} = (P_{\rm{tot}} - a{P^{*}_{\mathrm{R}}}) / N_{\rm{t}}$.
\end{Proposition}
{\it{Proof:}} By substituting $P_{\rm{B}} \!=\! (P_{\rm{tot}} \!-\! a P_{\rm{R}}) / N_{\rm{t}}$ into ${G}_{\mathrm{B}}$, 
we denote the channel gain of RDARS w.r.t. $P_{\mathrm{R}}$ as $f(P_{\mathrm{R}}) \!=\! \widetilde{f}^2(P_{\rm{R}})$, where  $\widetilde{f}(P_{\rm{R}})=\kappa_{\mathrm{br}} \sqrt {{P_{\rm{tot}}} - {aP_{\mathrm{R}}}}\sqrt {{N_{\rm{u}}}} ( {N - a} ) + {\kappa_{\mathrm{r}}}\sqrt {{P_{\mathrm{R}}}{N_{\rm{u}}}a} $. The first-order derivative of $\widetilde{f}(P_{\mathrm{R}})$ is given by ${\partial \widetilde{f}({P_{\mathrm{R}}})}/{{\partial P_{\mathrm{R}}}} \!=\! \mathsf{g}(P_{\rm{R}})/(2\sqrt P_{\mathrm{R}}{\sqrt {{P_{\rm{tot}}} - aP_{\mathrm{R}}}})$, where ${c_1} \!=\! {\kappa_{\mathrm{r}}}\sqrt {{N_{\rm{u}}}a}$, ${c_2} \!=\! {\kappa_{\mathrm{br}}}({N - a} )\sqrt {{N_{\rm{u}}}}$, and $\mathsf{g}(P_{\rm{R}}) \!=\! {{c_1}}\sqrt {{P_{\rm{tot}}} - aP_{\mathrm{R}}} - c_2 a \sqrt{P_{\mathrm{R}}}$. It is seen that the first-order derivative of $\mathsf{g}(P_{\rm{R}})$ is given by $\partial \mathsf{g}(P_{\rm{R}})/{{\partial P_{\mathrm{R}}}} = -(c_1 a) / (2{\sqrt {{P_{\rm{tot}}} - aP_{\mathrm{R}}}}) - (c_2a)/ (2\sqrt{P_{\mathrm{R}}})$. When $0<a\leq N$, we have $\partial \mathsf{g}(P_{\rm{R}})/{{\partial P_{\mathrm{R}}}} \leq 0$. When ${\partial f({P_{\mathrm{R}}})}/{{\partial P_{\mathrm{R}}}} =0$, the optimal transmit power budget for a transmit element is presented in \eqref{equ: P_R P_B in Propo 1}. Thus, the proof is completed.
~$\hfill\blacksquare$

By substituting ${P^{*}_{\mathrm{R}}}$ into \eqref{equ: rate for single UE ignore k}, 
the channel gain w.r.t. $a$ for the single UE is 
\begin{equation} \label{equ: channel gain RDARS} 
    f_{\rm{R}}(a)\!\!=\!\!\underbrace{{\kappa^{2}_{\mathrm{b}}}{P_{\rm{tot}}}{N_{\rm{u}}}{{( {N - a} )}^2}}_{\text{reflection gain}} \!\! +\!\! \underbrace{{\kappa^{2}_{\mathrm{r}}}{P_{\rm{tot}}}{N_{\rm{u}}}} _{\text{distributed gain}},  0\!\! < \!a \!\!\leq N.
\end{equation}
Then, the maximum rate w.r.t. $a$ is given by 
\begin{equation} \label{equ: Rate w.r.t. a} 
   \widetilde{{R}}_{\textrm{s}}(a)= \log_2 (1 + \frac{f_{\rm{R}}(a)}{\sigma^2} ).
\end{equation}
It is observed from \eqref{equ: Rate w.r.t. a} that the maximum rate of RDARS-aided system is determined by $a$,  which motivates the following proposition. 
\begin{Proposition}\label{propo: optimal a TDMA} 
The optimal number of RDARS transmit elements is $a_{\rm{T}}^* = 1$ when $1 \leq N < \widetilde{c}$ and $a_{\rm{T}}^* = 0$ when $N \geq \widetilde{c}$, with $\widetilde{c} = \frac{1}{2\kappa^{2}_{\mathrm{b}}} + \frac{1}{2}$.
\end{Proposition}
{\it{Proof:}} 
Since $f_{\rm{R}}(a)$ in \eqref{equ: channel gain RDARS} monotonically decreases when $0 < a \leq N$, the maximum rate is achieved at $a = 1$ when $0 < a \leq N$.
Then, we compare the rates for $a = 1$ and $a = 0$. 
Specifically, the channel gain at $a = 0$ is denoted by $f_{3}(N) = P_{\rm{tot}}\kappa^{2}_{\mathrm{br}}N_{\rm{u}} N^2$. 
By denoting $g_1 = f_{\rm{R}}(1) - f_3(N)$, it is seen that $ g_1 > 0$ holds when $N < {(1+\kappa^{2}_{\mathrm{br}}) / (2\kappa^{2}_{\mathrm{br}} )}$. Thus, the proof is completed.
~$\hfill\blacksquare$

Based on Proposition \ref{propo: optimal a TDMA} and $\omega_k = 1/K$, the maximum achievable WSR for the TDMA scheme is given by
\begin{align}\label{equ: max WSR in TDMA}
\widetilde{R}^{*} \!\!= \!\!
\begin{cases}
\!\!\! \frac{1}{K} \!\! \sum\limits_{k=1}^{K}{\! \log_2 (\! 1 \!\!+\!\! \frac{{\kappa^{2}_{\mathrm{r}, k}}{P_{\rm{tot}}}{N_{\rm{u}}}( {{\kappa^{2}_{\mathrm{b}}}{( {N \!-\! 1} )^2} \!+\! 1} )}{{\sigma^2}} \!)}, \! 1 \! \leq \! N \!<\! \widetilde{c}, \\ 
\!\!\! \frac{1}{K} \!\! \sum\limits_{k=1}^{K} \! \log_2 (\! 1\!\!+ \!\!\frac{P_{\rm{tot}}\kappa^{2}_{\mathrm{br},k}N_{\rm{u}} N^2}{{\sigma^2}} \!),\!  N \! \geq \! \widetilde{c}.
\end{cases}
\end{align}

Proposition \ref{propo: optimal a TDMA} shows that the RDARS-aided system always outperforms its DAS counterpart. In addition, the RDARS-aided system with the optimal $a$ is superior to its RIS-aided system counterpart with $1 \leq N < \widetilde{c}$. In the following, the performance comparison for different architectures is discussed in detail.

Let $g_2(N) = f_{\rm{R}}(N)-f_3(N)$,
and we have $g_2(N) > 0$ when $1 \leq N < \bar{c}$. This indicates the RDARS-aided system with $1 \leq a \leq N$ always outperforms the RIS-aided system when $1 \leq N  < \bar{c}$ with $\bar{c} = 1/{\kappa_{\rm{b}}}$. 
In addition, the channel gain of RDARS monotonically decreases with $a$, and thus only $1$ RDARS transmit element is required to achieve the maximum achievable rate in \eqref{equ: Rate w.r.t. a}.
To be specific, the reflection gain of RDARS decreases with $a$ due to the reduced number of passive elements. It is observed from \eqref{equ: P_R P_B in Propo 1} that the power budget allocated to RDARS transmit elements decreases with $a$ since the total power budget is limited. This makes the distributed gain independent of $a$. Thus, further increasing the number of RDARS transmit elements results in WSR loss, rather than performance improvement. Moreover, the placement position of the single RDARS transmit element is arbitrary under the LoS channel in this case. This is because the path loss from each RDARS element to UE is identical.
In fact, the threshold number of RDARS elements $\bar{c}$ is sufficiently large and hard to achieve in the practical scenario, thus guaranteeing the superiority of RDARS with/without mode switching. The above result demonstrates the inherent benefits of RDARS architecture.

On the other hand,
when $\bar{c} < N < \tilde{c}$, the RDARS-aided system with $1 \leq a \leq a_{\rm{T,th}}$ outperforms the RIS-aided system with $a_{\rm{T, th}} = N-\sqrt{N^2 - 1/\kappa^{2}_{\mathrm{b}}}$. 
It is observed that the number of RDARS elements ensuring that the RDARS outperforms RIS counterparts is independent of the links between RDARS and UEs.
Besides, the boundary of the total number of RDARS elements $\widetilde{c}$ only depends on the path loss of the BS-RDARS link.
When the path loss of BS-RDARS link is severe, the boundary $\widetilde{c}$ becomes sufficiently large since more reflection gain is required to compensate for the high path loss. 
In this case, the reflection gain is comparable to the distributed gain due to the sufficient number of passive elements.
This indicates that the extra distributed gain provided by deploying more RDARS transmit elements cannot compensate for the performance degradation when $a > a_{\rm{T, th}}$, only leading to higher power consumption.
Similarly, when $N > \widetilde{c}$, the reflection gain dominates the performance, where it is preferable to set all elements to work in reflection mode.

\begin{figure}[t] 
 \centering
 \includegraphics[width=0.4\textwidth]{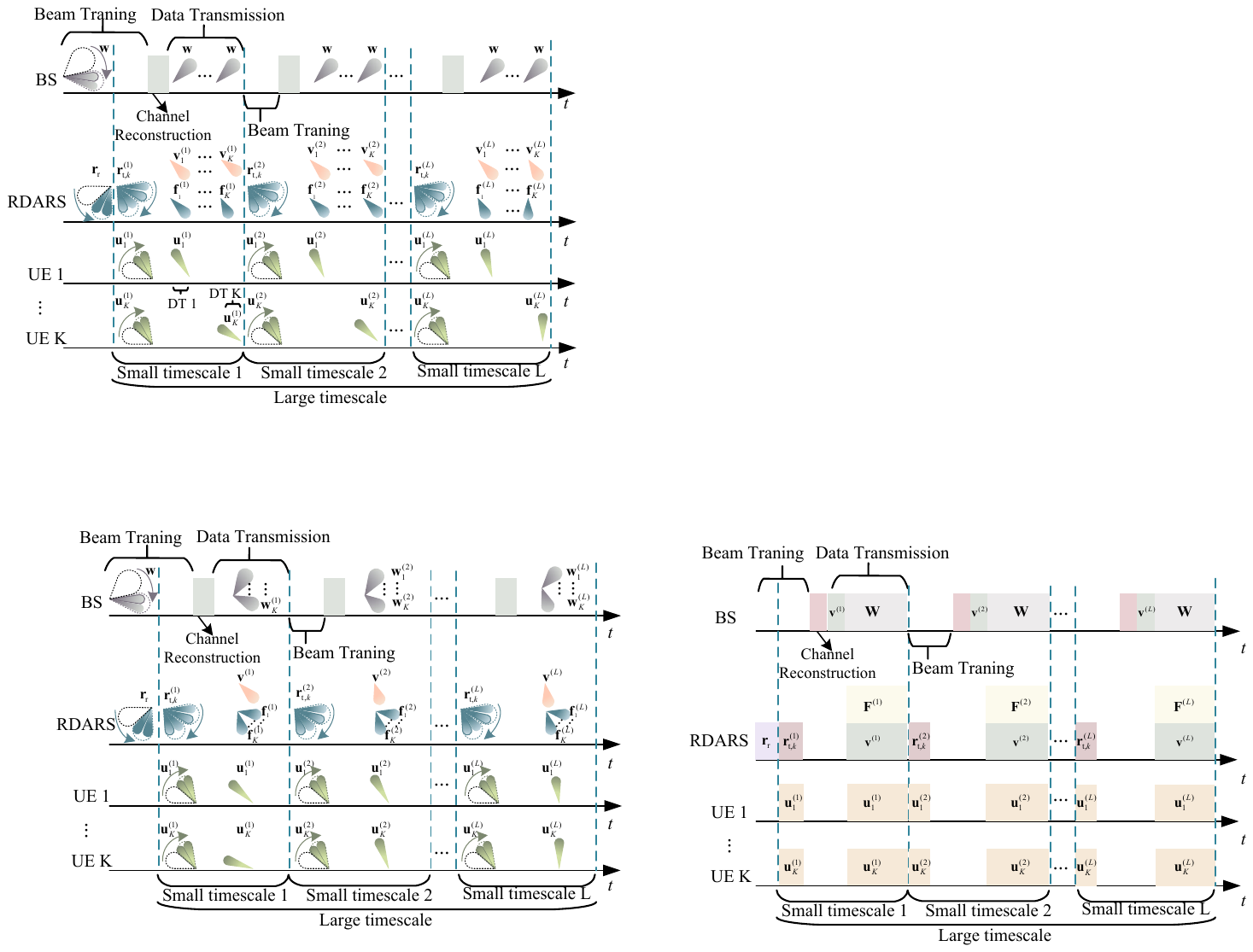}
 \vspace{-10pt}
 \caption{An illustration of the RCB-based BT and beamforming designs exploiting the two-timescale property for the TDMA scheme.}
 \label{fig: BT+data transmission in TDMA}
 \vspace{-20pt}
\end{figure}

Overall, the joint BT and beamforming design algorithm for the TDMA scheme is summarized as follows, which is illustrated in Fig. \ref{fig: BT+data transmission in TDMA}. 
The optimal beam pair $\{\mathbf{W}^{*}, \mathbf{F}^{*}, \mathbf{U}^{*}, \mathbf{\Phi}^{*}\}$ is obtained by solving problem \eqref{pro: single user fixed A, P_B, P_R}, followed by the determination of the power budget allocation matrix  $\mathbf{P}_{\mathrm{B}}^*$, $\mathbf{P}_{\mathrm{R}}^*$ and the mode switching matrix $\mathbf{A}^{*}$ via Propositions \ref{propo: optimal power allocation} and \ref{propo: optimal a TDMA}.

\subsubsection{SDMA} \label{subsec: SDMA}
For the SDMA scheme, it is challenging to directly obtain the optimal beamforming pairs for problem \eqref{pro: sum rate based on codebook} given $\bf{A}$, $\mathbf{P}_{\mathrm{B}}$ and $\mathbf{P}_{\mathrm{R}}$ due to the significant number of beam combinations. Moreover, there simultaneously exist multiple beams for UEs, where the shared BS-RDARS link leads to high IUI at the BS side. 
Besides, RDARS transmit elements generate the multi-beam to serve the UEs simultaneously, where the IUI at the RDARS side is also inevitable. 
To solve problem \eqref{pro: sum rate based on codebook}, it is transformed into two subproblems to maximize the weighted sum of SINR by mitigating the IUI at the BS and RDARS sides, respectively. 

Considering the IUI caused by BS antennas, each UE has the identical AoD from the BS to RDARS due to the common channel between the BS and RDARS. Therefore, the transmit beamforming vector for each UE to achieve the maximum channel gain is identical, i.e., $ \widetilde{\mathbf{w}}^{*}_{1} = \widetilde{\mathbf{w}}^{*}_{2} = \cdots = \widetilde{\mathbf{w}}^{*}_{K}$. Thus, the optimal codeword $\widetilde{\mathbf{b}}^{*}_{k}$ chosen from $\mathcal{C}_B$ for each UE is also the same one when the BT is conducted, thus yielding IUI at the BS side. 
For given $\mathbf{F}$, $\mathbf{A}$, $\mathbf{\Phi}$, $\mathbf{U}$, $\mathbf{P}_{\mathrm{B}}$ and $\mathbf{P}_{\mathrm{R}}$, problem \eqref{pro: sum rate based on codebook} is approximately reduced to 
\begin{equation}\label{pro: sum SINR at BS}
 \mathop {\max }\limits_{\{ \mathbf{w}_k\}} {\sum\limits_{k=1}^K \!\! \frac{ |\kappa_{{\rm{br}}, k}{{\bf{a}}^{\mathrm{H}}({{N_{\rm{t}}}}, \bar{\upsilon})} {\bf{w}}_k|^2 }{{ \sum\nolimits_{j=1, j \neq k}^K \!|\kappa_{{\rm{br}}, k} {{\bf{a}}^{\mathrm{H}}({{N_{\rm{t}}}}, \bar{\upsilon})} {\bf{w}}_j}|^2 \!+\! \sigma^2}}  
 \;\;\textrm{s.t.} \; \eqref{con: w CB}. 
\end{equation}
To suppress the IUI from the BS, the alternative codeword selection scheme is applied based on the channel gain. Specifically, the channel gain set $\mathcal{B} = \{\left[\mathcal{B}\right]_{\tau_i}, i =1,\cdots, N_{t}\}$ is obtained according to the codewords in $\mathcal{C}_{\mathrm{B}}$, where $\tau_i$ and $\left[\mathcal{B}\right]_{\tau_i}$ denote the index of the codeword in $\mathcal{C}_{\mathrm{B}}$ and the $i$-th largest channel gain, respectively, and $\mathcal{B}$ is sorted in a descending order. Let $\mathcal{C}_{t}$ denote the codeword set that will be allocated at the $t$-th round. At the first round $t=1$, the codeword set to be allocated is $\mathcal{C}_{1}$, where $\mathcal{C}_{1} = \mathcal{C}_{\mathrm{B}}$. Then, the codeword $\mathbf{b}_{\tau_1}$ that achieves the maximum channel gain is assigned to the UE with the maximum path loss. The alternative codeword set for the remaining UEs becomes $\mathcal{C}_{2}$, where $\mathcal{C}_{2} = \{\mathbf{b}_{\tau_i}, i = 2,\cdots, N_{\rm{t}}\}$. The codeword $\mathbf{b}_{\tau_2}$ is allocated to the UE with the sub-maximum path loss. By repeating the above steps until all beams are assigned, each beam can be uniquely assigned to the corresponding UE. As such, the optimized beamforming vectors at BS, ${\bf{w}}^{*}_k$, to problem \eqref{pro: sum SINR at BS} can be obtained.
With the IUI at the RDARS side, given $\mathbf{W}$, $\mathbf{A}$, $\mathbf{\Phi}$, $\mathbf{U}$, $\mathbf{P}_{\mathrm{B}}$ and $\mathbf{P}_{\mathrm{R}}$, problem \eqref{pro: sum rate based on codebook} is reduced to 
\begin{equation}\label{pro: sum SINR at RDARS}
 \mathop {\max }\limits_{\{ \mathbf{f}_k\}}
  {\sum\limits_{k=1}^K \!\frac{ |\kappa_{{\rm{r}}, k} {\widetilde{{\bf{a}}} ^{\mathrm{H}}}(N, \widetilde{\upsilon}_{k},\upsilon_{k}) {\bf{f}}_k|^2 }{{ \sum\limits_{j=1, j \neq k}^K \!\!\!\! |\kappa_{{\rm{r}}, k} {\widetilde{{\bf{a}}} ^{\mathrm{H}}}(N, \widetilde{\upsilon}_{k},\upsilon_{k}) {\bf{f}}_j}|^2 \!+\! \sigma^2}}  
 \;\;\;\textrm{s.t.}\;\;\eqref{con: f CB}.
\end{equation}

In the following, we first design an orthogonal connected RCB associated with the UE distribution to mitigate the IUI caused by the multi-beam from RDARS transmit elements.  
Since the connected RCB depends on the number and placement positions of RDARS transmit elements, the mode switching matrix can be obtained by designing the orthogonal transmit RCB for RDARS transmit elements, as shown in the following propositions.
\begin{Proposition} \label{propo: optimal placement SDMA}
For the SDMA scheme, to guarantee the orthogonality of each two codewords, the locations of RDARS transmit elements should satisfy 
\begin{align}\label{con: position on z, y}
z_m & = z_1 + q^{*}(m-1)d_{\rm{R}}, \; 0 \leq m < a_z, \\
 y_n & = y_1 + p^{*}(n-1)d_{\rm{R}}, \; 0 \leq n < a_y, 
\end{align} 
with 
\begin{align}\label{con: optimal placement in propo 2}
    & q^{*} \in \{q\in \mathbb{Z}^{+}| 1 \leq q \leq \lfloor\frac{N_{\rm{z}}-1}{a_{\rm{z}}-1}\rfloor , \gcd(q, a_{\rm{z}}) = 1 \},\\
    & p^{*} \in \{p\in \mathbb{Z}^{+}| 1 \leq p \leq \lfloor\frac{N_{\rm{y}}-1}{a_{\rm{y}}-1}\rfloor, \gcd(p, a_{\rm{y}}) = 1 \},
\end{align}
where $m \in \mathbb{Z}^+, n \in \mathbb{Z}^+$, $\gcd(q, a_{\rm{z}}) = 1$ indicates that $q$ and $a_{\rm{z}}$ are coprime, and $a_{\rm{z}}$ and $a_{\rm{y}}$ denote the numbers of connected elements along $z$ and $y$ directions, respectively.
\end{Proposition}
{\it{Proof:}} Please refer to Appendix A.
~$\hfill\blacksquare$

Proposition \ref{propo: optimal placement SDMA} shows that the optimal placement positions of RDARS transmit elements depend on the number of transmit elements, which may not be unique. Besides, the orthogonality of the codeword can be ensured when the phase difference caused by the propagation distance is compensated for the LoS channel. It is also worth mentioning that the optimal placement yields a parallel distribution for transmit elements along the $\sf{y}$ and $\sf{z}$ directions. 

Furthermore, for the connected codebook, the beam coverage of each codeword is determined by the number of transmit elements, given by $\mathcal{C}\mathcal{V}\left({\mathbf{c}}_z\right) = 2/a_{\rm{z}}$ and $\mathcal{C}\mathcal{V}\left({\mathbf{c}}_y\right) = 2/a_{\rm{y}}$, where $\mathcal{C}\mathcal{V}(\cdot)$ denotes the coverage of a beam in the angle domain.
In addition, the optimal beam angle for each UE chosen from the connected RCB should be aligned with the spatial direction corresponding to the virtual AoD of the RDARS-UE channel for each UE. Therefore, the choice of the optimal beam for the UE should mitigate the IUI at the RDARS side. 
Then, we have the following proposition.
\begin{Proposition} \label{propo: optimal a}
For the SDMA scheme, the set of the optimal number of transmit elements is given by $a_{\mathrm{S}}^{*} = a^{*}_z \times a^{*}_y$, where 
\begin{align}
 a^{*}_z & \in \{a_{\rm{z}} \in \mathbb{Z}^{+} | \max\{ K, 2/\widetilde{K}\} \leq a_{\rm{z}} \leq N_{\rm{z}}\}, \label{equ: optimal a in SDMA, z axis}\\
 a^{*}_y  &\in \{ a_{\rm{y}} \in \mathbb{Z}^{+} | \max\{ K,2/\bar{K}\}  \leq a_{\rm{y}} \leq N_{\rm{y}}\}\label{equ: optimal a in SDMA, y axis},
\end{align}
with $\widetilde{K} = \min \{\left|\widetilde{\upsilon}_m -\widetilde{\upsilon}_n\right|\}$, $\bar{K} =\min \{\left|\upsilon_{m} - \upsilon_{n}\right|\}$, and $m \neq n, \forall m,n\in \mathcal{K}$. The minimum of transmit elements is $a^{*}_{\rm{S, th}} = \operatorname{max}\{K, 2/\widetilde{K}\} \times \operatorname{max}\{K, 2/\bar{K}\}$.
\end{Proposition}
{\it{Proof:}} Please refer to Appendix B.
~$\hfill\blacksquare$

\begin{figure}[tb] 
 \centering
 \includegraphics[width=0.38\textwidth]{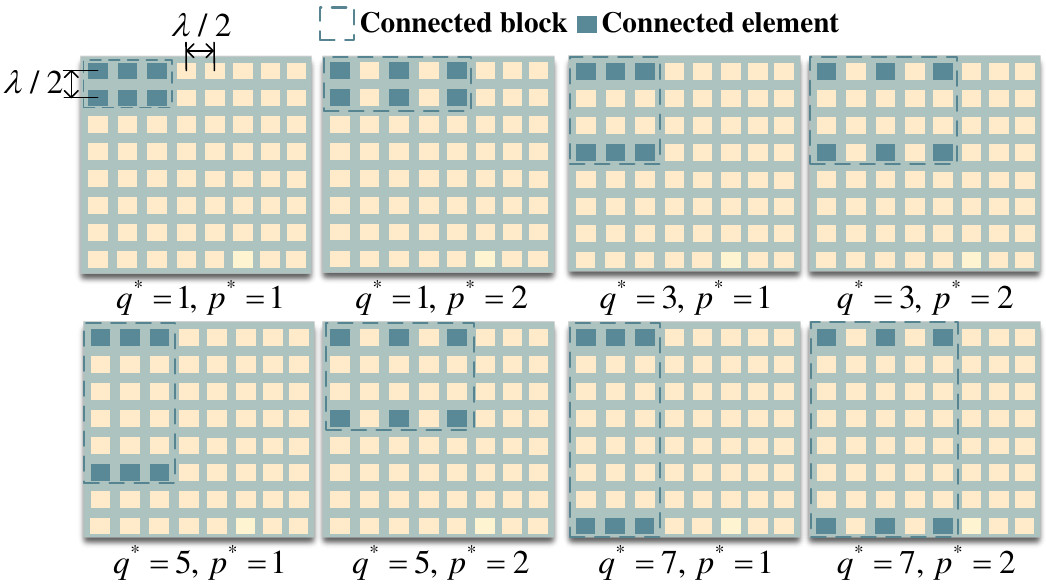}
 \vspace{-10pt}
 \caption{An example of the placement positions of connected elements for a $8 \times 8$ RDARS.}
 \label{fig: an example of connected elements}
 \vspace{-15pt}
\end{figure}

A closer look at \eqref{equ: optimal a in SDMA, z axis} and \eqref{equ: optimal a in SDMA, y axis} shows that the threshold for the number of RDARS transmit elements is determined by the UE distribution. When the UE distribution is very dense, a significantly large number of transmit elements are required to achieve the high spatial resolution, and vice versa. Thus, it is preferable to allocate the elements more than the threshold as transmit elements for the SDMA scheme. In addition, the optimal spatial placement positions of transmit elements in \eqref{con: optimal placement in propo 2} provide considerable flexibility in adjusting the RDARS element spacing from a hardware design perspective. It is observed that the optimal spatial placements of transmit elements can always be found for given $d_{\mathrm{R}}$. Besides, multiple optimal choices of placement position provide greater robustness at the hardware level, as illustrated in Fig. \ref{fig: an example of connected elements}.

With Propositions \ref{propo: optimal placement SDMA} and \ref{propo: optimal a}, the channel gain of the $k$-th UE is given by
\begin{align}\label{equ: max_gain for F}
\widetilde{G}_{k} 
& = |\widetilde{{\bf{a}}}^{\mathrm{H}}(a,\widetilde{\theta}_{z, k},\widetilde{\theta}_{\mathrm{zy}, k})\mathbf{x}_{\mathrm{R}}|^2 \nonumber \\
& = |\frac{1}{\sqrt{a}}\widetilde{{\bf{a}}}^{\mathrm{H}}(a,\widetilde{\upsilon}_k,{\upsilon}_k ) \widetilde{{\bf{a}}}(a,-1+\frac{2i-1}{a_{\rm{z}}},-1+\frac{2j-1}{a_{\rm{y}}} ) |^2 \nonumber \\
& =|\widetilde{G}_{k_{\rm{z}}} \cdot G_{k_{\rm{y}}} /{\sqrt{a}}|^2\mathop \to \limits^{a.s.} a,
\end{align}
where $i, j \in \mathbb{Z}^{+}$, $0 \leq i \leq a_{\rm{z}}$ and $0 \leq j \leq a_{\rm{y}}$.
It is observed that $i$ and $j$ always exist for the $k$-th UE to ensure that $\widetilde{G}_{k} \to a$ when $a_{\rm{z}}, a_{\rm{y}} \to \infty$. 
The result shows that the interference minimization problem is equivalent to the channel gain maximization problem, which is also equivalent to the maximum WSR problem under the condition $a_{\rm{z}}, a_{\rm{y}} \to \infty$. In addition, the maximum channel gain can also be achieved when all the virtual AoDs of the RDARS-UE channels are located in the set of spatial directions formed by the connected RCB. However, the latter condition may not be satisfied due to the UE mobility and channel variations. Therefore, $\mathbf{f}_{k}^{*}$ chosen from the connected RCB for RDARS transmit elements is in general a high-quality solution to problem (\ref{pro: sum rate based on codebook}). Furthermore, the equation in (\ref{equ: max_gain for F}) demonstrates that the codeword achieving the maximum channel gain along each axis is independent of each other. This implies that the resolution with respect to the specific axis direction of the RCB only depends on the density of the UE distribution in the corresponding direction. 

Given $\mathbf{W}^{*}$, $\mathbf{F}^{*}$, $\mathbf{A}^{*}$, and $\mathbf{U}^{*}$, problem \eqref{pro: sum rate based on codebook} is reduced to
\begin{align}\label{pro: sum rate based on codebook in SDMA water filling}
 \mathop {\max }\limits_{\mathbf{\Phi},\mathbf{P}_{\mathrm{B}}, \mathbf{P}_{\mathrm{R}}}
 \;\;{\sum\nolimits_{k = 1}^K {\omega_k\log_2(1+ \widetilde{\gamma}_k) }}
 \;\;\textrm{s.t.}\;\;\eqref{con: Phi CB}, \eqref{con: P},
\end{align}
where $\widetilde{\gamma}_k = \frac{|\mathbf{u}^{*\mathrm{H}}_k \mathbf{H}_{k} \mathbf{f}^{*}_{\mathrm{R},k}|^2 }{\sum_{ i \neq k}^{K}|\mathbf{u}^{*\mathrm{H}}_k \mathbf{H}_{k} \mathbf{f}^{*}_{\mathrm{R},i }|^2 + \sigma^2 }$. 

Given $\mathbf{P}_{\mathrm{B}}$ and $\mathbf{P}_{\mathrm{R}}$, the phase shift matrix $\mathbf{\Phi}$ is optimized via the one-dimensional search over $\mathcal{C}_{\mathrm{R},2D}$. Given optimized $\mathbf{\Phi}$, the power budget allocation matrices $\mathbf{P}_{\mathrm{B}}$ and $\mathbf{P}_{\mathrm{R}}$ are optimized by the water-filling method at the presence of the IUI \cite{water_filling}. Then, the passive beamforming vector and power control are alternately optimized until convergence is achieved.

\begin{figure}[t] 
 \centering
 \includegraphics[width=0.4\textwidth]{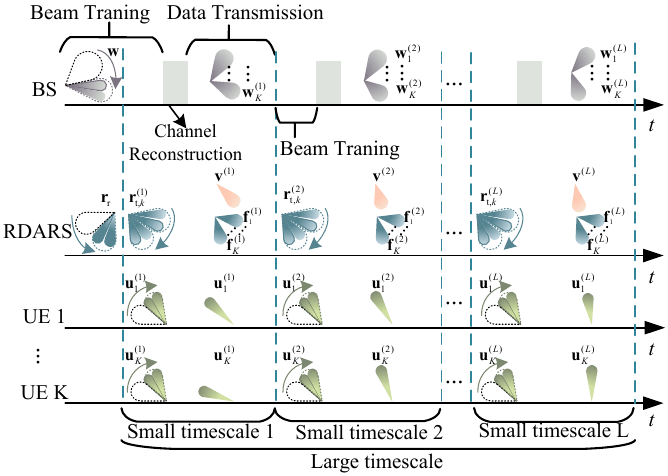}
 \vspace{-10pt}
 \caption{An illustration of the RCB-based BT and beamforming designs exploiting the two-timescale property for the SDMA scheme.}
 \label{fig: BT+data transmission in SDMA}
 \vspace{-15pt}
\end{figure}

Overall, the joint BT and beamforming design algorithm for the SDMA scheme is summarized as follows, which is illustrated in Fig. \ref{fig: BT+data transmission in SDMA}. With the equal power allocation, the optimal beam pair $\{\mathbf{W}^{*}, \mathbf{F}^{*}\}$ is obtained by solving problem \eqref{pro: sum SINR at BS} and \eqref{pro: sum SINR at RDARS}. Besides, $\mathbf{U}^{*}$ is obtained by the MRC beamforming, followed by the determination of the mode switching matrix $\mathbf{A}^{*}$ via Propositions \ref{propo: optimal placement SDMA} and \ref{propo: optimal a}. Then, the optimal $\mathbf{\Phi}^{*}$ and the optimal power budget allocation matrices $\mathbf{P}_{\mathrm{B}}^*$ and $\mathbf{P}_{\mathrm{R}}^*$ are obtained by solving problem \eqref{pro: sum rate based on codebook in SDMA water filling}. 

\section{Simulation Results}\label{sec: simulation}
In this section, we provide simulation results to validate the effectiveness of the proposed schemes and to draw useful insights for RDARS-aided mmWave MU-MIMO systems.
The number of antennas at the BS is $N_{\rm{t}} = 64$, and that of elements of the RDARS is $N = 128$. The number of UEs is $K = 3$.
The BS and RDARS are located at (0, 0, 15) meter (m) and (10, 0, 15) m, respectively, and the UEs are randomly distributed within a radius of 5 m centered at (10, 50, 2) m.
The path loss exponents of both the BS-RDARS and RDARS-UE links are set as 2.0 and 2.2, respectively. 
The signal attenuation at a reference distance of 1 m is set as 60.4 dB.
Moreover, the power of the noise is set as -80 dBm.

For comparison, the following architectures are considered: \textbf{1) RDARS:} The RDARS-aided system is considered; \textbf{2) DAS:} The DAS is considered, where transmit beamforming vectors are designed and the number of distributed antennas is set as $a_{\rm{d}} = a$; \textbf{3) RIS:} For the benchmark architecture of RIS, the number of RIS elements is set as $\widetilde{N} = N$.

\subsection{Comparisons Among Different Schemes}
\begin{figure*}[!t]
	\centering
	\begin{minipage}[b]{0.325\textwidth}
        \vspace{-10pt}
		\includegraphics[width=\textwidth]{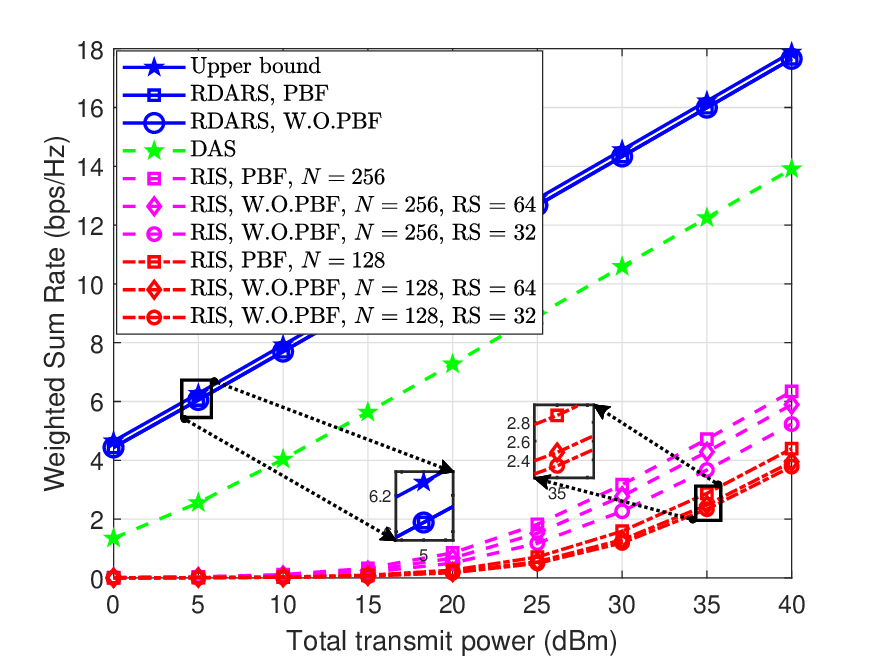}
		\vspace{-20pt}
		\caption{WSR versus the total transmit power for the TDMA scheme.}
        \vspace{-10pt}
		\label{fig: BT schemes in TDMA}
	\end{minipage}
	\begin{minipage}[b]{0.325\textwidth}
        \vspace{-10pt}
		\includegraphics[width=\textwidth]{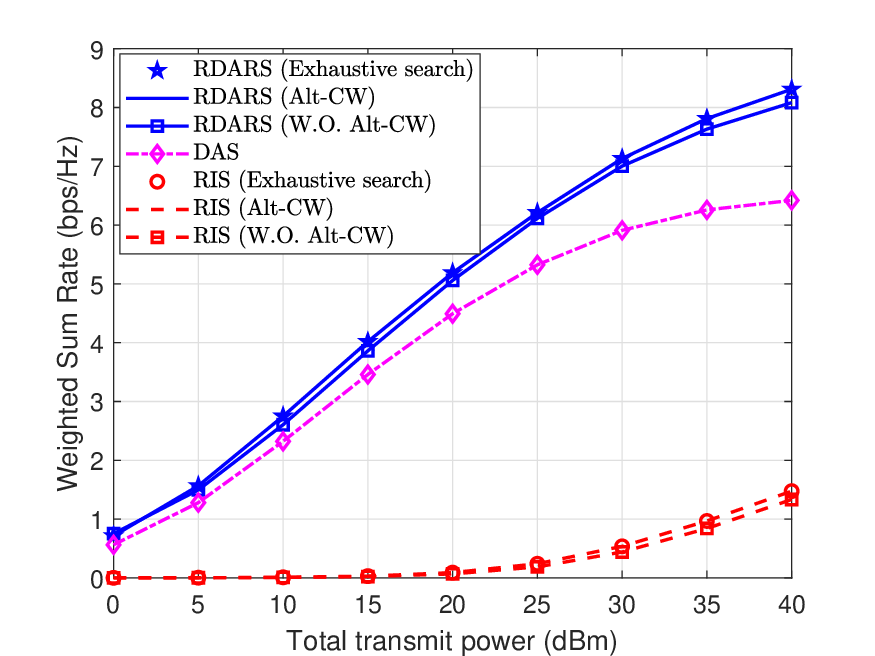}
		\vspace{-20pt}
		\caption{Impacts of the alternative codewords scheme on the WSR for the SDMA scheme.}
        \vspace{-10pt}
	 \label{fig: Alter codewords in SDMA in data transmission}
	\end{minipage}
	\begin{minipage}[b]{0.325\textwidth}
        \vspace{-10pt}
		\includegraphics[width=\textwidth]{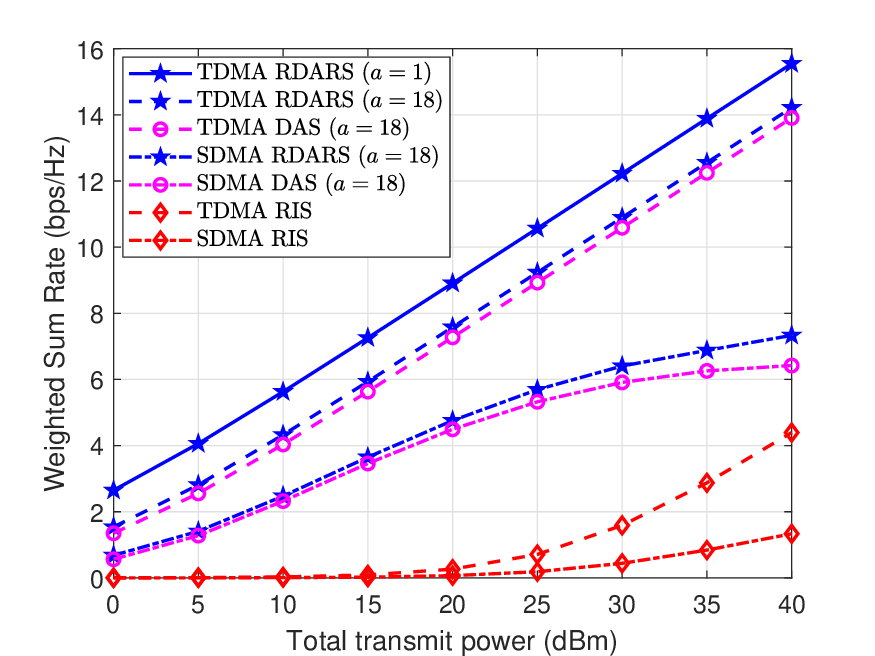}
		\vspace{-20pt}
		\caption{WSR versus the total transmit power.}
        \vspace{-4pt}
 \label{fig: RDARS VS DAS VS RIS}
	\end{minipage}
\end{figure*}

In this subsection, we compare different beamforming schemes in terms of the WSR for the TDMA scheme in Fig. \ref{fig: BT schemes in TDMA}. 
The following schemes are considered for comparison: \textbf{1) PBF:}
The active and passive beamforming for the BS, RDARS passive elements, RDARS transmit elements, and UEs are obtained via the exhaustive search; \textbf{2) W.O.PBF:} The beamforming design is carried out based on the proposed scheme, where the passive beamforming vector is reconstructed according to Section \ref{sec: TDMA}; \textbf{3) Upper bound:} The maximum WSR is calculated for the TDMA scheme according to (\ref{equ: max WSR in TDMA}).
It is observed that the WSR increases as $P_{\rm{tot}}$ increases for both the PBF and W.O.PBF schemes due to the increasing beamforming gain, and the proposed scheme incurs a very slight performance loss compared to the upper bound. This is because the obtained optimal pair can achieve beam alignment with the corresponding channel. 
The gap between the upper bound and PBF for RDARS is constant as $P_{\rm{tot}}$ increases due to the finite and fixed resolution for BS antennas, RDARS passive elements, and transmit elements.
Moreover, the performance of the proposed scheme, i.e., W.O.PBF, is comparable with that of the PBF scheme, which indicates that the proposed scheme can achieve nearly optimal performance with low overhead.  

In Fig. \ref{fig: Alter codewords in SDMA in data transmission}, we compare the WSR of the different beamforming designs at the BS side, i.e., the exhaustive search, the alternative codeword scheme (\textbf{Alt-CW}) and the scheme without the alternative codeword (\textbf{W.O. Alt-CW}) versus the total transmit power for the SDMA scheme, where the number of RDARS elements is $N = 512$. 
It is observed that the performance of the Alt-CW scheme yields similar performance to the exhaustive search. This is expected since the codeword achieving the large channel gain is allocated to the UE suffering from severe path loss, so as to maximize the WSR.
It is also observed that the Alt-CW scheme provides a maximum of 2 dB gain compared with the W.O. Alt-CW scheme, and the gap between these two schemes increases with the total transmit power. 
This is because the IUI caused by beam conflicts exaggerates the performance loss with the increasing transmit power.
By contrast, the proposed beamforming design provides a high-quality solution to the transmit beamforming vectors at the BS side with low computational overhead.

\subsection{Comparisons Among Different Architectures}\label{subsec: Impact of the architecture}
In Fig. \ref{fig: RDARS VS DAS VS RIS}, 
we plot the WSR of the different architectures for TDMA and SDMA schemes versus the total transmit power.
It is observed that the RDARS-aided system for the TDMA scheme with $a = 1$ achieves the maximum WSR. This is because the IUI is eliminated for the TDMA scheme, and the power budget is optimally allocated for RDARS elements.
Besides, it is observed that the RDARS-aided system outperforms its DAS and RIS-aided system counterparts in given MA schemes. The reason is that the selection gain is provided by the dynamic mode selection.
Additionally, the curves for the RDARS-aided system and DAS for the TDMA scheme are linear compared with those for the SDMA scheme. 
This is because the IUI cannot be eliminated for the SDMA scheme despite the applications of alternative codewords, the orthogonal RCBs, and the optimized number and placement positions of transmit elements. 
Specifically, the performance gap between the RDARS-aided system for the TDMA scheme and that for the SDMA scheme increases as the total transmit power. 
The reason is that the IUI is more severe with the increasing total transmit power, and finally the curves flatten out for the SDMA scheme. 
This demonstrates the superiority of RDARS and verifies the effectiveness of beamforming designs in different MA schemes.
\vspace{-8pt}
\subsection{Impact of the Number of BS Antennas}
\begin{figure}[t] 
 \centering
 \includegraphics[width=0.325\textwidth]{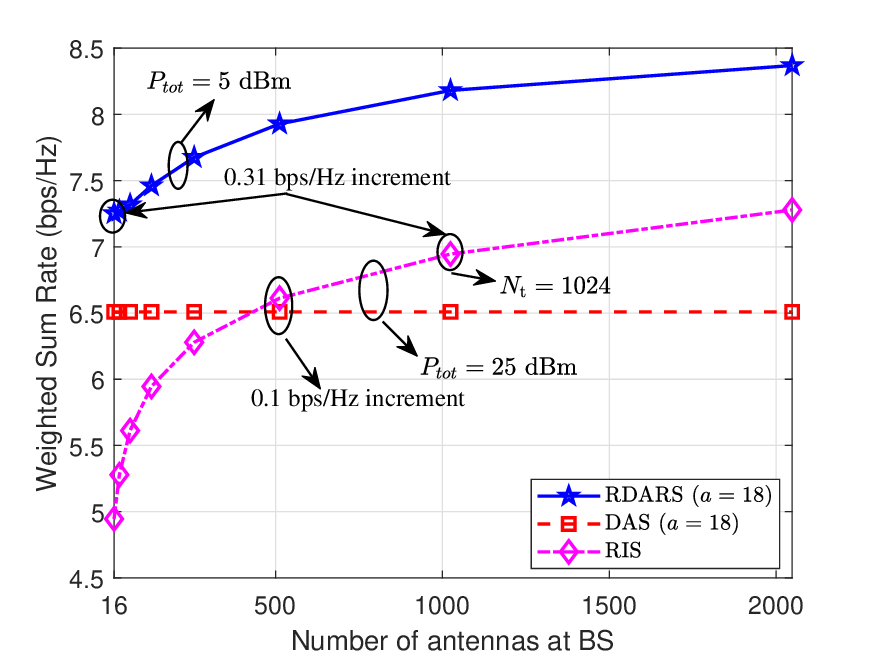}
 \vspace{-5pt}
 \caption{WSR versus the number of the BS antennas for the SDMA scheme.}
 \label{fig: impact of the number of antennas at BS}
 \vspace{-15pt}
 \end{figure}

In Fig. \ref{fig: impact of the number of antennas at BS}, 
we plot the WSR of different architectures versus the number of BS antennas, by considering the different total transmit power for the SDMA scheme with $N = 128$.
It is observed from Fig. \ref{fig: impact of the number of antennas at BS} that the RDARS-aided system outperforms the DAS system for different values of $N_{\rm{t}}$. Moreover, the RDARS-aided system with only a small number of transmit elements, i.e., $a = 18$, and $16$ BS antennas at $P_{\rm{tot}} = 5$ dBm outperforms the RIS-aided system with 1024 antennas by 0.31 bps/Hz at $P_{\rm{tot}} = 25$ dBm. 
Besides, the RDARS-aided system with $a = 18$ and zero passive elements $P_{\rm{tot}} = 5$ dBm has a comparable performance than the RIS-aided system with 500 antennas and 128 passive elements at $P_{\rm{tot}} = 25$ dBm. 
The reason is that a small number of transmit elements is required to compensate for the multiplicative fading.
Meanwhile, increasing the number of passive elements is more energy-efficient and cost-effective than further adding more BS antennas.
This demonstrates the great potential to reduce the number of RF chains and energy consumption for RDARS-aided systems, highlighting the benefits of incorporating RDARS into massive MIMO due to the low cost and energy consumption of passive elements.

\subsection{Impact of the Codebook Design}
\begin{figure*}[!t]
	\centering
	\begin{minipage}[b]{0.325\textwidth}
		\includegraphics[width=\textwidth]{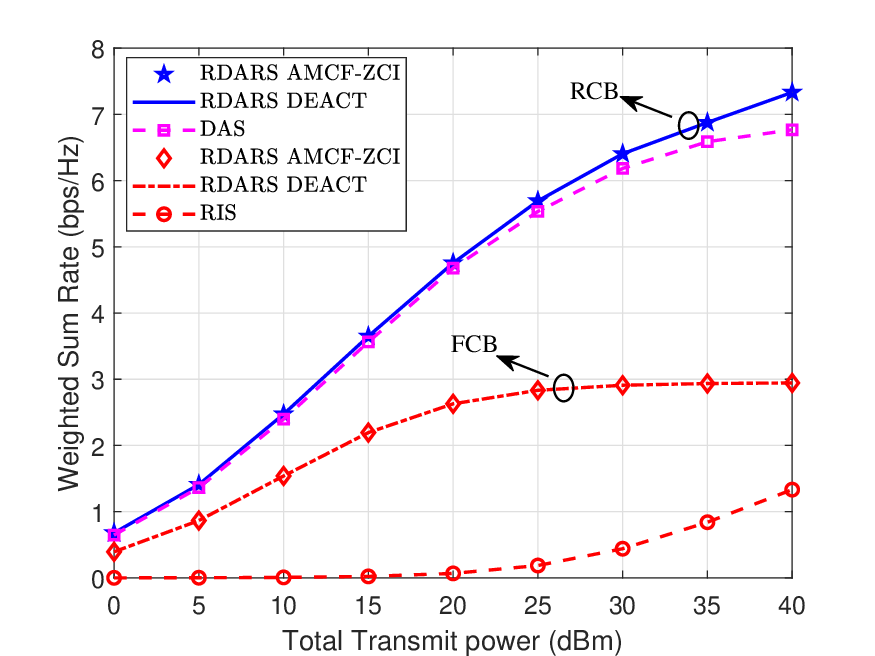}
		\vspace{-20pt}
		\caption{WSR versus the total transmit power for the SDMA scheme.}
        \vspace{-10pt}
		\label{fig: reconfigurable Codebook schemes}
	\end{minipage}
	\begin{minipage}[b]{0.325\textwidth}
		\includegraphics[width=\textwidth]{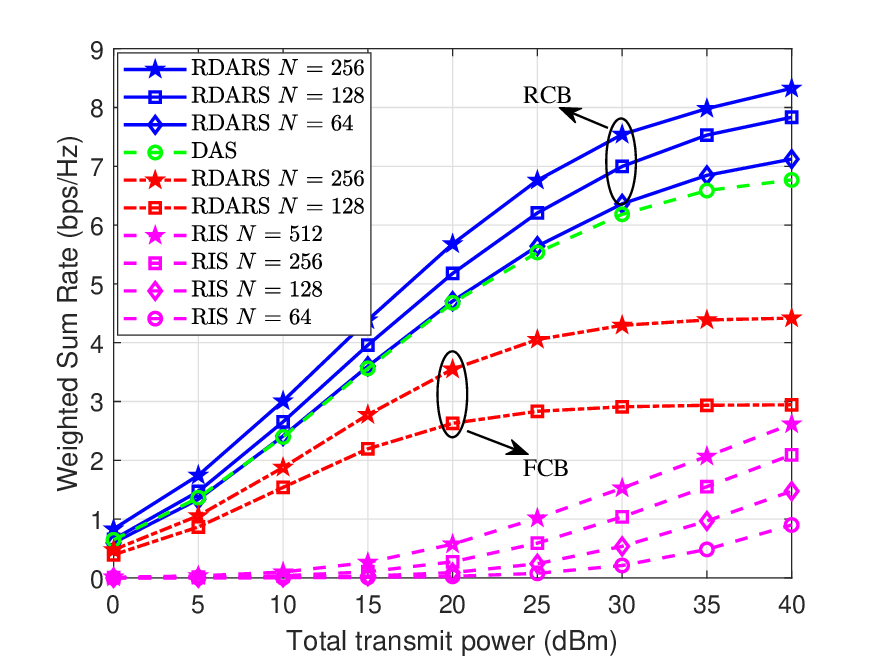}
		\vspace{-20pt}
		       \caption{WSR versus the total transmit power for the SDMA scheme under different $N$.}
        \vspace{-10pt}
		\label{fig: impact of number of passive elements}
	\end{minipage}
	\begin{minipage}[b]{0.325\textwidth}
		\includegraphics[width=\textwidth]{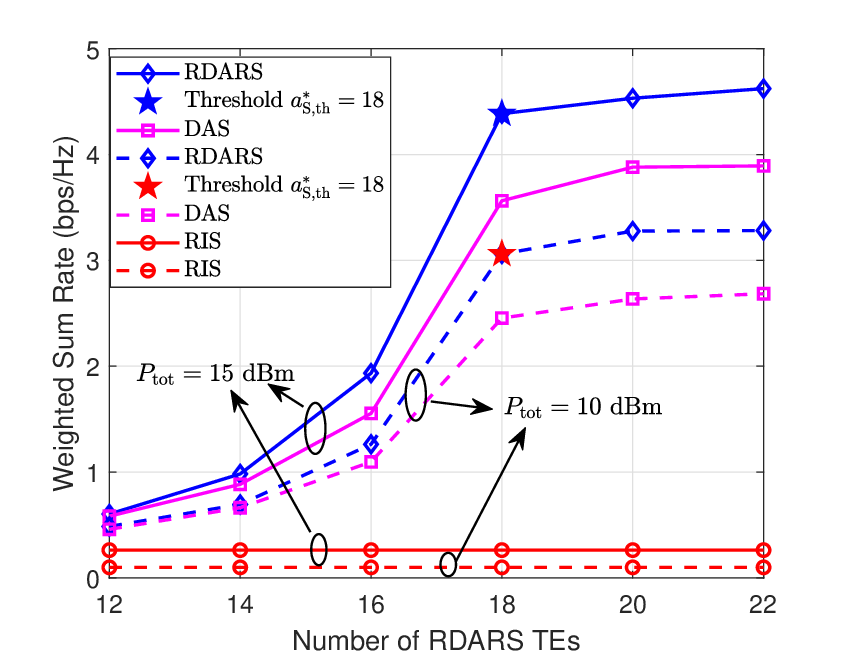}
		\vspace{-20pt}
		\caption{WSR versus the number of TEs for the SDMA scheme under different $P_{\rm{tot}}$.}
        \vspace{-10pt}
 \label{fig: impact of number for transmit elements in SDMA}
	\end{minipage}
\end{figure*}

In Fig. \ref{fig: reconfigurable Codebook schemes}, 
we plot the WSR of different codebooks versus the total transmit power for the SDMA scheme.
Two codebook design schemes at the bottom layer of the AMCF-ZCI and DEACT codebooks are considered for comparison: \textbf{1) RCB:} The codebook is generated according to ${\bf{A}}^{*}$ and Section \ref{sec: codebook design}; \textbf{2) FCB:} The conventional fixed codebook is applied \cite{ChenhaoQi_HS, ZhenyuXiao_HS}.
\begin{figure}[t] 
 \centering
 \includegraphics[width=0.325\textwidth]{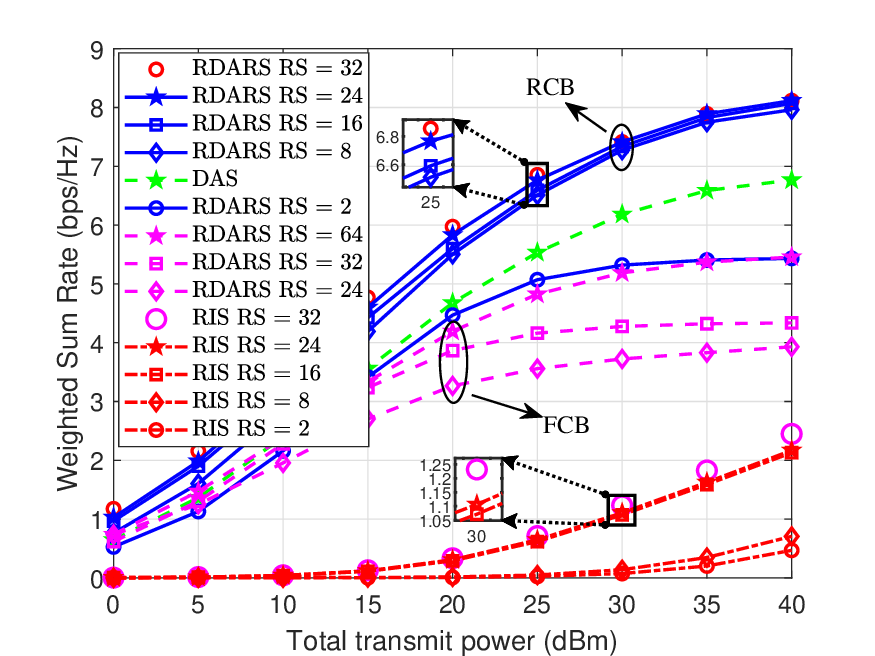}
 \caption{WSR versus the total transmit power for the SDMA scheme under different RCB resolutions.}
\label{fig: impact of resolution for passive elements}
 \end{figure}
It is observed that the two curves of the AMCF-ZCI and DEACT yields a quite similar WSR with the increasing total transmit power. This is expected since the same codeword at the bottom layers is applied to the two codebook schemes. 
Moreover, the AMCF-ZCI and DEACT approaches with the RCB outperform those with the FCB. This is because the RCB compensates for the phase difference caused by the flexible reconfiguration of RDARS elements. 
Additionally, it is seen that both of the curves of RCB and FCB schemes tend to flatten out as the total transmit power increases. To be specific, the RCB outperforms the FCB by 3 bps/Hz at $P_{\rm{tot}} = 25$, under which the performance curve with the FCB begins to flatten. This gap will continue to increase until it reaches a sufficiently large power.
The reason is that the fixed phase difference for codewords causes intolerable phase misalignment, thus resulting in high IUI. 
Moreover, It is observed that the performance of RDARS-aided system is worse than that of the DAS without applying the RCB, which demonstrates the importance of the RCB to unleash the full potential of RDARS in maximizing the WSR.
\vspace{-5pt}
\subsection{Impact of the Number of RDARS Elements}
In Fig. \ref{fig: impact of number of passive elements}, 
we plot the WSR of different architectures versus the total transmit power, by considering different numbers of passive elements and codebook types with $a = 18$.
It is observed that the performance increases with the number of passive elements for RDARS and RIS-aided systems. The reason is that a larger passive beamforming gain can be achieved as the number of reflection elements increases. 
Besides, the RDARS-aided system significantly outperforms the RIS-aided system and DAS, especially for a large number of passive elements. 
As the total transmit power increases, such a performance gap becomes more evident since RDARS is capable of well mitigating the IUI by leveraging the connected elements and RCB codebook design. 
In addition, it is observed that the performance improvement with FCB is more pronounced compared with that with RCB in the RDARS-aided system. This is because a larger number of elements is required to compensate well for phase misalignment with FCB.
Again, it is observed that the performance of RDARS-aided systems with $N = 256$ is even worse than that of the DAS if the FCB is applied, which further verifies the benefits of RCB.

In Fig. \ref{fig: impact of number for transmit elements in SDMA}, we investigate the IUI mitigation capability for RDARS transmit elements, by plotting the WSR of the RDARS-aided system versus the number of transmit elements with $N = 512$. 
In this simulation, the set of the virtual vertical AoDs and horizontal AoDs from RDARS transmit elements to UEs are set as $\{-0.96875, -0.59375, -0.21875\}$ and $\{-0.84375, -0.03125, 0.78125\}$, respectively. 
According to Proposition \ref{propo: optimal a}, under this UE deployment, the minimum number of transmit elements can be found as $a^{*}_{\rm{S}, th}=18$. The simulation results in Fig. \ref{fig: impact of number for transmit elements in SDMA} are consistent with the analytical result in Proposition \ref{propo: optimal a}. 
Specifically, it is observed that the WSR drops sharply when the number of RDARS transmit elements is below the threshold. This is because an insufficient number of RDARS transmit elements leads to beam conflict for the UEs, thus incurring performance degradation, especially when the distributed gain dominates the performance.
\subsection{Impact of the Codebook Resolution for the Passive Elements}
In Fig. \ref{fig: impact of resolution for passive elements}, 
we plot the WSR of different architectures with different resolutions and different codebook types versus the total transmit power at $N = 256$, where RS refers to the codebook resolution for passive elements along each axis.
It is observed that the performance of RDARS and RIS-aided systems increases as the codebook resolution improves. Note that the curves of ${\rm{RS}} = 32$ and those of ${\rm{RS}} = 24$ are very similar since the resolution is sufficiently large in this case.
In addition, it is observed that the performance of RDARS-aided systems is inferior to that of the DAS when the resolution is small. e.g., ${\rm{RS}} = 2$. The reason is that the passive beamforming gain is limited by the low resolution.
Moreover, the FCB scheme experiences a significant performance enhancement as the resolution increases, since it can provide a higher spatial diversity to mitigate the influence of phase misalignment.
\section{Conclusion} \label{sec: conclusion}
In this paper, we investigated a RDARS-aided downlink multi-user MIMO system, by designing the RCB-based BT and beamforming for the TDMA and SDMA schemes. Under the TDMA scheme, the optimal number of RDARS transmit elements and power allocation are derived in closed form. Based on the closed-form WSR, the superiority of RDARS was demonstrated in terms of WSR when the total number of RDARS elements was moderate. Under the SDMA scheme, the IUI from RDARS transmit elements was mitigated for the WSR maximization when the number of transmit elements was more than a threshold and RDARS transmit elements were located in a parallel type. Simulation results unveiled the RCB design outperforms the fixed codebook due to its great adaptability. Moreover, the alternative codeword scheme achieved an approximate performance compared with the exhaustive search for minimizing the IUI at the BS. The results in this paper showed the effectiveness of our schemes and illustrated that integrating them into RDARS-aided MU-MIMO systems is a promising approach for improving performance in 6G mmWave communications.

\vspace{-3pt}
\section*{Appendix A: Proof of Proposition \ref{propo: optimal placement SDMA}} \label{appendix: proof optimal placement SDMA}
\vspace{-5pt}
The array response vector of RDARS transmit elements is $\breve{{ {\bf{a}}}}(\mathsf{z}\otimes \mathsf{y}, N, \theta_{z,k},\theta_{\mathrm{y},k})\!\! =\!\! \widehat{\bf{a}}_(\mathsf{z}, {N_{\rm{z}}},\theta_{z,k}) \otimes \widehat{\bf{a}}(\mathsf{y},{N_{\rm{y}}}, \theta_{y,k})$. We denote $\widetilde{\theta}_{\mathrm{zy},k} = \widetilde{\upsilon}_{k}\upsilon_{k}$.
The beam pattern of the transmit array is $G(\mathsf{z},\mathsf{y},\mathbf{f}_k,\widetilde{\theta}_{\mathrm{zy},k})= |\breve{{ {\bf{a}}}}^{\rm{H}}(\mathsf{z}\otimes \mathsf{y}, N,  \widetilde{\upsilon}_{k},\upsilon_{k})\mathbf{f}_k|^2$.
Note that $\mathbf{f}_k$ is chosen from the bottom layer of $C^{\mathrm{c}}_{\rm{R}}$, denoted as $\mathcal{F}_{\textrm{sub}}= \{\mathbf{f}(\xi_f), f = 1,\cdots, a\}$, which is to be designed by determining $a$, with $\xi_f$ being the spatial direction of the $f$-th codeword. To achieve the maximum channel gain, the phase of the chosen codeword for each UE should be closest to that of the related channel. Therefore, the best codeword of $\mathcal{F}_{\textrm{sub}}$ is given by $\mathbf{f}(\xi_f) = \frac{1}{\sqrt{a}}\breve{{\bf{a}}}(\mathsf{z} \otimes \mathsf{y},a, \xi_{f_{\rm{z}}}, \xi_{f_{\rm{y}}})$, where $\xi_f = \xi_{f_{\rm{z}}}\xi_{f_{\rm{y}}}$, $\breve{{\bf{a}}}(\mathsf{z} \otimes \mathsf{y},a, \xi_{f_{\rm{z}}}, \xi_{f_{\rm{y}}}) =\widehat{{\bf{a}}}_{}(\mathsf{z} ,a_{\rm{z}}, \xi_{f_{\rm{z}}}) \otimes \widehat{{\bf{a}}}(\mathsf{y} ,a_{\rm{y}},\xi_{f_{\rm{y}}})$, $\xi_{f_{\rm{z}}} = -1+\frac{2f_{\rm{z}}-1}{a_{\rm{z}}}$, $\xi_{f_{\rm{y}}} = -1+\frac{2f_{\rm{y}}-1}{a_{\rm{y}}}$ and $f = f_{\rm{z}} \times f_{\rm{y}}$.
For ease of exposition, the arbitrary two codewords $\mathbf{f}_{\rm{z}}(\xi_{f_{zi}})$ and $\mathbf{f}_{\rm{z}}(\xi_{f_{zj}})$ along the $z$-axis are considered, where $\mathbf{f}_{\rm{z}}(\xi_{f_{zi}}) = 1/\sqrt{a_{\rm{z}}} \widehat{{\bf{a}}}(\mathsf{z}, {a_{\rm{z}}}, \xi_{f_{zi}})$ and $\mathbf{f}_{\rm{y}}(\xi_{f_{yi}}) = 1/\sqrt{a_{\rm{y}}} \widehat{{\bf{a}}}(\mathsf{y}, {a_{\rm{y}}}, \xi_{f_{yi}})$.

Considering the placement positions of transmit elements, the multi-beam interference for $\mathbf{f}_{\rm{z}}(\xi_{f_{zi}})$ and $\mathbf{f}_{\rm{z}}(\xi_{f_{zj}})$ is $  \bar{I}_{ij}
= 1/a_{\rm{z}}\sum\nolimits_{n=1}^{a_{\rm{z}}} {e^{\jmath\frac{2\pi \cdot 2\left({f_{zi} - f_{zj}}\right)}{\lambda \cdot a_{\rm{z}}}z_{n}}} \nonumber
= 1/a_{\rm{z}}\sum\nolimits_{m=1}^{q} {(1 - e^{\jmath\frac{ 2\pi \cdot  2\left({f_{zi}} - {f_{zj}}\right)}{\lambda \cdot a_{\rm{z}}} \ell_m d_m} )/(1 - e^{\jmath\frac{ 2\pi \cdot  2\left({f_{zi} - f_{zj}}\right)}{\lambda \cdot a_{\rm{z}}} d_m} } )$,
where $q$ denotes the number of element sets, and $\ell_m$ and $d_m$ denote the number of transmit elements and element spacing in the $m$-th element set, respectively. To guarantee that $\bar{I}_{ij} = 0$ for given $i$ and $j$, we have $\frac{ 2\pi \cdot  2|{f_{zi}} - {f_{zj}}|}{\lambda \cdot a_{\rm{z}}} \ell_m d_m = 2k_{m}\pi$ and $\frac{ 2\pi \cdot  2|{f_{zi}} - {f_{zj}}|}{\lambda \cdot a_{\rm{z}}} d_m \neq 2k^{'}_{m}\pi$, where $k_m, k^{'}_{m} \in \mathbb{Z}$. Therefore, the element spacing in the $m$-th element set should satisfy
\begin{align} \label{equ: d_m}
d_m = k_{m}\frac{\lambda \cdot a_{\rm{z}}}{2 \cdot \ell_m |{f_{zi}} - {f_{zj}}|}, d_m \neq k^{'}_{m}\frac{\lambda \cdot a_{\rm{z}}}{2 \cdot |{f_{zi}} - {f_{zj}}|}.
\end{align}

Note that \eqref{equ: d_m} holds if and only if $\ell_m = a_{\rm{z}}$ and $q = 1$. Therefore, (\ref{equ: d_m}) is transformed into $2/\lambda |{f_{zi}} - {f_{zj}}| d_{\rm{z}} = k$ and $2/\lambda |{f_{zi}} - {f_{zj}}| d_{\rm{z}}/a_{\rm{z}} \neq k^{'}$, where $d_{\rm{z}} = nd_{\rm{R}}$. Without loss of generality, we assume that $d_{\rm{R}} = \lambda/2$. Therefore, the element spacing along the $z$-axis should satisfy $n|{f_{zi}} - {f_{zj}}|=k$ and $\frac{n}{a_{\rm{z}}}|{f_{zi}} - {f_{zj}}| \neq k^{'}$.
It is observed that $|{f_{zi}} - {f_{zj}}| \in \mathsf{K}$, where $\mathsf{K} = \{\widetilde{k}\in \mathbb{Z}^{+}| 1\leq \widetilde{k} \leq a_{\rm{z}} - 1 \}$. Therefore, we have $n \in \{n\in \mathbb{Z}^{+}| 1 \leq n \leq N_{\rm{z}}-1 , \gcd(n, a_{\rm{z}}) = 1 \}$, where $\gcd(n, a_{\rm{z}}) = 1$ denotes $n$ and $a_{\rm{z}}$ are coprime. 
Furthermore, the channel gains in the two coordinate directions are independent of each other if the orthogonal codeword is applied. 
The constraint for the connected element along the $y$-axis is $m \in \{m\in \mathbb{Z}^{+}| 1 \leq m \leq N_{\rm{y}}-1 , \gcd(m, a_{\rm{y}}) = 1 \}$.
Therefore, we have
 $\frac{1}{a_{\rm{z}}}\widehat{{\bf{a}}}^{\mathrm{H}}(\mathsf{z},a_{\rm{z}},\xi_{i_{\mathrm{z}}}){{\bf{a}}}(\mathsf{z} ,a_{\rm{z}},\xi_{j_{\mathrm{z}}}) = 
\delta_{ij}$ and 
$\frac{1}{a_{\rm{y}}}\widehat{{\bf{a}}}^{\mathrm{H}}_{}(\mathsf{y} ,a_{\rm{y}},\xi_{i_{\mathrm{y}}})\widehat{{\bf{a}}}_{}(\mathsf{y},a_{\rm{y}}, \xi_{j_{\mathrm{y}}}) = 
\delta_{ij}$,
where $z_m = z_1 + (m-1)qd_{\mathrm{R}}$ with 
\begin{align}
q \in \{q\in \mathbb{Z}^{+}| 1 \leq q \leq N_{\rm{z}}-1 , \gcd(q, a_{\rm{z}}) = 1 \}, \label{con: position on z}
\end{align} 
and $y_n = y_1 + (n-1)pd_{\mathrm{R}}$ with 
\begin{align}
p \in \{p\in \mathbb{Z}^{+}| 1 \leq p \leq N_{\rm{y}}-1 , \gcd(p, a_{\rm{y}}) = 1 \}. \label{con: position on y}
\end{align}

Furthermore, it follows that ${\mathbf{f}}^{\mathrm{H}}(\xi_i){\mathbf{f}}(\xi_j) = 1$ when $i = j$ and ${\mathbf{f}}^{\mathrm{H}}(\xi_i){\mathbf{f}}(\xi_j) = 1$ otherwise. The optimal $q$ and $p$ are given by $q^* = {k\lambda}/{(2d_{\mathrm{R}}})$ and $p^* = {k^{'}\lambda}/{(2d_{\mathrm{R}}})$, where $k$ and $k^{'}$ are positive integers which ensure that $q^*$ and $p^*$ satisfy ($\ref{con: position on z}$) and ($\ref{con: position on y}$), respectively. Considering the array aperture, we have $q(a_{\rm{z}} - 1)d_{\mathrm{R}} \leq (N_{\rm{z}}-1)d_{\mathrm{R}}$ and $p(a_{\rm{y}} - 1)d_{\mathrm{R}} \leq (N_{\rm{y}}-1)d_{\mathrm{R}}$.  Therefore, given $d_{\mathrm{R}} = \lambda/2$, the optimal $q$ and $p$ are given by $q^{*} \in \{q\in \mathbb{Z}^{+}| 1 \leq q \leq \lfloor\frac{N_{\rm{z}}-1}{a_{\rm{z}}-1}\rfloor , \gcd(q, a_{\rm{z}}) = 1 \}$ and $p^{*} \in \{p\in \mathbb{Z}^{+}| 1 \leq p \leq \lfloor\frac{N_{\rm{y}}-1}{a_{\rm{y}}-1}\rfloor, \gcd(p, a_{\rm{y}}) = 1 \}$, respectively.
\vspace{-5pt}
\section*{Appendix B: Proof of Proposition \ref{propo: optimal a}} \label{appendix: proof optimal number SDMA}
Let $L_k = \sum\nolimits_{i=1,i\neq k}^{K} {L_{k,i}}$ denote the multi-interference for the k-th UE, where ${L_{k,i}}$ denotes the interference between the $k$-th UE and the $i$-th UE. It is observed that $L_k = 0$ holds when ${L_{k,i}} = 0$ satisfies for each $i$.
For the RDARS-UE channel, ${L_{k,i}}$ is given by  
\begin{align} \label{equ: inter for k-th UE}
     {L_{k,i}}\!\!  = &|\breve{\mathbf{a}}^{\rm{H}} (\mathsf{z}\otimes \mathsf{y}, a, \theta_{z,k},\theta_{y,k} )\breve{\mathbf{a}} (\mathsf{z}\otimes \mathsf{y},a, \theta_{z,j},\theta_{y,j} ) |^2 \nonumber \\
      = & L_{\mathrm{z}, k} L_{\mathrm{y}, k},
\end{align}
where $\theta_{\rm{z}} \in [-1, 1]$ and $\theta_{y} \in [-1, 1]$ respectively denote the spatial directions of the beam along the $z$ axis and $y$-axis. Let $L_{\mathsf{z}, k}$ and $L_{\mathsf{y}, k}$ respectively denote the interference along the $z$ axis and $y$-axis. It is observed from (\ref{equ: inter for k-th UE}) that the interference along the $z$-axis is independent of the other along the $y$-axis. Therefore, in the following, the proof of the optimal number of transmit elements along the $z$-axis is first shown, and that with respect to the y-axis can be similarly obtained, which is omitted for brevity.

Specifically, we denote the virtual AoD related to the RDARS-UE channel for UE $k$ is $\widetilde{\varphi}_k$. Therefore, based on Proposition \ref{propo: optimal placement SDMA}, the optimal spatial direction of the beam is $\theta^{*}_k = \mathop{\arg\min}\nolimits_{\theta_k} |\theta_k - \widetilde{\varphi}_k|$. 
Then, to guarantee the uniqueness of $\theta^{*}_k$ for $k$-th UE, the coverage of ${\mathbf{c}}_z$ along the $z$-axis direction satisfies $\mathcal{C}\mathcal{V}\left({\mathbf{c}}_z\right) = {2}/{a_{\rm{z}}} \leq \left|\widetilde{\varphi}_m- \widetilde{\varphi}_n \right|$.
Meanwhile, we have $a_{\rm{z}} \geq K$ and $a_{\rm{y}} \geq K$ to ensure a sufficient number of data streams. Note that $\widetilde{\varphi}_k = \widetilde{\varepsilon}_k$ holds according to (\ref{equ: Channel BR}). 
As a result, the number of transmit elements along the $z$-axis satisfies \begin{equation}
\max\{ K, 2/\widetilde{K} \} \leq a_{\rm{z}} \leq N_{\rm{z}}, \; a_{\rm{z}} \in \mathbb{Z}^{+},\label{range: a_z}
\end{equation}
with $\widetilde{K} = \min \{\left|\widetilde{\upsilon}_m -\widetilde{\upsilon}_n\right|\}, m \neq n, \forall m,n\in \mathcal{K}$. Similarly, the number of transmit elements along the $y$-axis satisfies 
\begin{equation}
\max\{ K, 2/\bar{K}\} \leq a_{\rm{y}} \leq N_{\rm{y}}, \; a_{\rm{y}} \in \mathbb{Z}^{+}, \label{range: a_y}  
\end{equation}
with $\bar{K} =\min \{\left|\upsilon_{m} - \upsilon_{n}\right|\}, m \neq n, \forall m,n\in \mathcal{K}$. Therefore, the optimal set of transmit elements is $a_{{\rm{S}}}^{*} = a^{*}_{\rm{z}} \times a^{*}_{\rm{y}}$, where $a^{*}_{\rm{z}}$ and $a^{*}_{\rm{y}}$ satisfy (\ref{range: a_z}) and (\ref{range: a_y}), respectively.

\bibliographystyle{IEEEtran}
\vspace{-5pt}
\bibliography{IEEEabrv, Reference}

\begin{thebibliography}{10}
\providecommand{\url}[1]{#1}
\csname url@samestyle\endcsname
\providecommand{\newblock}{\relax}
\providecommand{\bibinfo}[2]{#2}
\providecommand{\BIBentrySTDinterwordspacing}{\spaceskip=0pt\relax}
\providecommand{\BIBentryALTinterwordstretchfactor}{4}
\providecommand{\BIBentryALTinterwordspacing}{\spaceskip=\fontdimen2\font plus
\BIBentryALTinterwordstretchfactor\fontdimen3\font minus
  \fontdimen4\font\relax}
\providecommand{\BIBforeignlanguage}[2]{{%
\expandafter\ifx\csname l@#1\endcsname\relax
\typeout{** WARNING: IEEEtran.bst: No hyphenation pattern has been}%
\typeout{** loaded for the language `#1'. Using the pattern for}%
\typeout{** the default language instead.}%
\else
\language=\csname l@#1\endcsname
\fi
#2}}
\providecommand{\BIBdecl}{\relax}
\BIBdecl

\bibitem{xue_Survey}
Q.~Xue \emph{et~al.}, ``A survey of beam management for {mmWave} and {THz}
  communications towards {6G},'' \emph{{IEEE} Commun. Surveys Tuts.}, vol.~26,
  no.~3, pp. 1520--1559, 3rd Quart., 2024.

\bibitem{NadezhdaChukhno_Survey}
N.~Chukhno \emph{et~al.}, ``Models, methods, and solutions for multicasting in
  {5G/6G} {mmWave} and sub-{THz} systems,'' \emph{{IEEE} Commun. Surveys
  Tuts.}, vol.~26, no.~1, pp. 119--159, 1st Quart., 2024.

\bibitem{Robert_DAS}
R.~W.~H. Jr. \emph{et~al.}, ``A current perspective on distributed antenna
  systems for the downlink of cellular systems,'' \emph{{IEEE} Commun. Mag.},
  vol.~51, no.~4, pp. 161--167, Apr. 2013.

\bibitem{Antonino_C_RAN}
A.~Masaracchia \emph{et~al.}, ``Digital twin for open {RAN:} {Toward}
  intelligent and resilient {6G} radio access networks,'' \emph{{IEEE} Commun.
  Mag.}, vol.~61, no.~11, pp. 112--118, Nov. 2023.

\bibitem{Smruti_C_RAN}
S.~R. Swain \emph{et~al.}, ``An {AI}-driven intelligent traffic management
  model for {6G} cloud radio access networks,'' \emph{{IEEE} Wireless Commun.
  Lett.}, vol.~12, no.~6, pp. 1056--1060, Jun. 2023.

\bibitem{haiquan_survey}
H.~Lu \emph{et~al.}, ``A tutorial on near-field {XL-MIMO} communications
  towards {6G},'' \emph{{IEEE} Commun. Surv. Tutorials}, vol.~26, no.~4, pp.
  2213--2257, 4th Quart., 2024.

\bibitem{Andr_CF}
A.~R. Flores \emph{et~al.}, ``Clustered cell-free multi-user multiple-antenna
  systems with rate-splitting: Precoder design and power allocation,''
  \emph{{IEEE} Trans. Commun.}, vol.~71, no.~10, pp. 5920--5934, Oct. 2023.

\bibitem{Ertugrul_RIS}
E.~Basar, ``Reconfigurable intelligent surface-based index modulation: {A} new
  beyond {{MIMO}} paradigm for {6G},'' \emph{{IEEE} Trans. Commun.}, vol.~68,
  no.~5, pp. 3187--3196, May 2020.

\bibitem{xue_DoubleRIS}
Q.~Xue \emph{et~al.}, ``Multi-user {mmWave} uplink communications based on
  collaborative double-{RIS}: {Joint} beamforming and power control,''
  \emph{{IEEE} Commun. Lett.}, vol.~27, no.~10, pp. 2702--2706, Oct. 2023.

\bibitem{haiquanlu_RIS}
H.~Lu \emph{et~al.}, ``Aerial intelligent reflecting surface: Joint placement
  and passive beamforming design with {3D} beam flattening,'' \emph{{IEEE}
  Trans. Wireless Commun.}, vol.~20, no.~7, pp. 4128--4143, Jul. 2021.

\bibitem{AhmedElzanaty_RIS}
A.~Elzanaty \emph{et~al.}, ``Reconfigurable intelligent surfaces for
  localization: Position and orientation error bounds,'' \emph{{IEEE} Trans.
  Signal Process.}, vol.~69, pp. 5386--5402, Aug. 2021.

\bibitem{Shuhaozeng_RIS}
S.~Zeng \emph{et~al.}, ``Reconfigurable intelligent surface {(RIS)} assisted
  wireless coverage extension: {RIS} orientation and location optimization,''
  \emph{{IEEE} Commun. Lett.}, vol.~25, no.~1, pp. 269--273, Jan. 2021.

\bibitem{RuizheLong_Multifading}
R.~Long \emph{et~al.}, ``Active reconfigurable intelligent surface-aided
  wireless communications,'' \emph{{IEEE} Trans. Wireless Commun.}, vol.~20,
  no.~8, pp. 4962--4975, Aug. 2021.

\bibitem{Peng_Power_constraint}
Q.~Peng \emph{et~al.}, ``Hybrid active-passive {IRS} assisted energy-efficient
  wireless communication,'' \emph{{IEEE} Commun. Lett.}, vol.~27, no.~8, pp.
  2202--2206, Jul. 2023.

\bibitem{Rafaela_HybridRIS}
R.~Schroeder \emph{et~al.}, ``Two-stage channel estimation for hybrid {RIS}
  assisted {{MIMO}} systems,'' \emph{{IEEE} Trans. Commun.}, vol.~70, no.~7,
  pp. 4793--4806, Jul. 2022.

\bibitem{peng_semiRIS}
Q.~Peng \emph{et~al.}, ``Semi-passive intelligent reflecting surface enabled
  sensing systems,'' \emph{{IEEE} Trans. Commun.}, vol.~72, no.~12, pp.
  7674--7688, Dec. 2024.

\bibitem{ChengzhiMa_ANewArchi}
C.~Ma \emph{et~al.}, ``Reconfigurable distributed antennas and reflecting
  surface ({RDARS}): {A} new architecture for wireless communications,''
  \emph{{IEEE} Trans. Commun.}, vol.~72, no.~10, pp. 6583--6598, Oct. 2024.

\bibitem{zhang_RDARS}
P.~Zhang \emph{et~al.}, ``Integrated sensing and communication with
  reconfigurable distributed antenna and reflecting surface: Joint beamforming
  and mode selection,'' \emph{arXiv preprint arXiv:2401.05182}, 2024.

\bibitem{Wang_RDARS}
J.~Wang \emph{et~al.}, ``Joint beamforming optimization and mode selection for
  {RDARS}-aided {{MIMO}} systems,'' \emph{{IEEE} Trans. Wireless Commun.},
  vol.~23, no.~11, pp. 17\,557--17\,572, Nov. 2024.

\bibitem{ChengzhiMa_RDARS}
C.~Ma \emph{et~al.}, ``{RDARS} empowered massive {{MIMO}} system:
  {Two}-timescale transceiver design with imperfect {CSI},'' \emph{{IEEE}
  Trans. Wireless Commun.}, vol.~23, pp. 18\,806--18\,821, Dec. 2024.

\bibitem{jintao_ISAC_RDARS}
J.~Wang \emph{et~al.}, ``Demo: {Reconfigurable} distributed antennas and
  reflecting surface ({RDARS})-aided integrated sensing and communication
  system,'' in \emph{Proc. {IEEE} ICCC}, 2023, pp. 1--2.

\bibitem{HuanHuang_TT}
H.~Huang \emph{et~al.}, ``Two-timescale-based beam training for {RIS}-aided
  millimeter-wave multi-user {MISO} systems,'' \emph{{IEEE} Trans. Veh.
  Technol.}, vol.~72, no.~9, pp. 11\,884--11\,897, Sep. 2023.

\bibitem{SungGeunYoon_Codebook}
S.~Yoon \emph{et~al.}, ``Improved hierarchical codebook-based channel
  estimation for mmwave massive {{MIMO}} systems,'' \emph{{IEEE} Wireless
  Commun. Lett.}, vol.~11, no.~10, pp. 2095--2099, Oct. 2022.

\bibitem{jiancheng_Codebook}
J.~An \emph{et~al.}, ``Codebook-based solutions for reconfigurable intelligent
  surfaces and their open challenges,'' \emph{{IEEE} Wireless Commun.},
  vol.~31, no.~2, pp. 134--141, Apr. 2024.

\bibitem{DeyouZhang_CB}
D.~Zhang \emph{et~al.}, ``Codebook-based training beam sequence design for
  millimeter-wave tracking systems,'' \emph{{IEEE} Trans. Wireless Commun.},
  vol.~18, no.~11, pp. 5333--5349, Nov. 2019.

\bibitem{JingheWang_Shutdown}
J.~Wang \emph{et~al.}, ``Hierarchical codebook-based beam training for
  {RIS}-assisted {mmWave} communication systems,'' \emph{{IEEE} Trans.
  Commun.}, vol.~71, no.~6, pp. 3650--3662, Jun. 2023.

\bibitem{ZhenyuXiao_HS}
Z.~Xiao \emph{et~al.}, ``Hierarchical codebook design for beamforming training
  in millimeter-wave communication,'' \emph{{IEEE} Trans. Wireless Commun.},
  vol.~15, no.~5, pp. 3380--3392, May 2016.

\bibitem{ChenhaoQi_HS}
C.~Qi \emph{et~al.}, ``Hierarchical codebook-based multiuser beam training for
  millimeter wave massive {{MIMO}},'' \emph{{IEEE} Trans. Wireless Commun.},
  vol.~19, no.~12, pp. 8142--8152, Dec. 2020.

\bibitem{YuLu_HS}
Y.~Lu \emph{et~al.}, ``Hierarchical beam training for extremely large-scale
  {{MIMO}:} {From} far-field to near-field,'' \emph{{IEEE} Trans. Commun.},
  vol.~72, no.~4, pp. 2247--2259, Apr. 2024.

\bibitem{ChangshengYou_Static}
C.~You \emph{et~al.}, ``Fast beam training for {IRS}-assisted multiuser
  communications,'' \emph{{IEEE} Wireless Commun. Lett.}, vol.~9, no.~11, pp.
  1845--1849, Nov. 2020.

\bibitem{PeilanWang_RIS_Passive}
P.~Wang \emph{et~al.}, ``Beam training and alignment for {RIS}-assisted
  millimeter-wave systems: State of the art and beyond,'' \emph{{IEEE} Wireless
  Commun.}, vol.~29, no.~6, pp. 64--71, May 2022.

\bibitem{ChenchengZhang_RIS_single_UE}
C.~Zhang \emph{et~al.}, ``Fast multibeam training for {RIS}-assisted millimeter
  wave massive {MIMO},'' \emph{{IEEE} Commun. Lett.}, vol.~28, no.~1, pp.
  168--172, Jan. 2024.

\bibitem{Yuhao_Chen_RIS_single_UE}
Y.~Chen and L.~Dai, ``Coded beam training for {RIS} assisted wireless
  communications,'' \emph{{IEEE} Trans. Wireless Commun.}, 2025, Early Access,
  doi: {10.1109/TWC.2025.3535160}.

\bibitem{Xing_Jia_Codebook_BF_MISO}
X.~Jia \emph{et~al.}, ``Environment-aware codebook for reconfigurable
  intelligent surface-aided {MISO} communications,'' \emph{{IEEE} Wireless
  Commun. Lett.}, vol.~12, no.~7, pp. 1174--1178, Jul. 2023.

\bibitem{Xu_Shi_Codebook_BF_single_UE}
X.~Shi \emph{et~al.}, ``Spatial-chirp codebook-based hierarchical beam training
  for extremely large-scale massive {MIMO},'' \emph{{IEEE} Trans. Wireless
  Commun.}, vol.~23, no.~4, pp. 2824--2838, Apr. 2024.

\bibitem{Baishuo_Lin_C_BF_MU}
B.~Lin \emph{et~al.}, ``Low-complexity two-timescale hybrid precoding for
  mmwave massive {MIMO:} {A} group-and-codebook based approach,'' \emph{{IEEE}
  Trans. Wireless Commun.}, vol.~23, no.~7, pp. 7263--7277, Jul. 2024.

\bibitem{NLOS_Ignore}
B.~Wang \emph{et~al.}, ``Spectrum and energy-efficient beamspace {MIMO-NOMA}
  for millimeter-wave communications using lens antenna array,'' \emph{{IEEE}
  J. Sel. Areas Commun.}, vol.~35, no.~10, pp. 2370--2382, Oct. 2017.

\bibitem{Xianghao_NLOS}
X.~Yu \emph{et~al.}, ``Coverage analysis for millimeter wave networks: The
  impact of directional antenna arrays,'' \emph{{IEEE} J. Sel. Areas Commun.},
  vol.~35, no.~7, pp. 1498--1512, Jul. 2017.

\bibitem{water_filling}
W.~Yu \emph{et~al.}, ``Iterative water-filling for gaussian vector
  multiple-access channels,'' \emph{{IEEE} Trans. Inf. Theory}, vol.~50, no.~1,
  pp. 145--152, Jan. 2004.

\end{thebibliography}

\end{document}